\documentclass[twocolumn]{aastex631}
\UseRawInputEncoding
\usepackage{color}
\usepackage{lineno}
\newcommand\scalemath[2]{\scalebox{#1}{\mbox{\ensuremath{\displaystyle #2}}}}

\shortauthors{Zhu et al.}

\begin{document}
	
\title{Theoretical Diagnostics for Narrow Line Regions of Active Galactic Nuclei}

\author[0000-0002-1333-147X]{Peixin Zhu}
\affiliation{Research School of Astronomy and Astrophysics, Australian National University, Australia}
\affiliation{ARC Centre of Excellence for All Sky Astrophysics in 3 Dimensions (ASTRO 3D), Australia}
\affiliation{Center for Astrophysics $|$ Harvard \& Smithsonian, 60 Garden Street, Cambridge, MA 02138, USA}

\author[0000-0001-8152-3943]{Lisa J. Kewley}
\affiliation{Research School of Astronomy and Astrophysics, Australian National University, Australia}
\affiliation{ARC Centre of Excellence for All Sky Astrophysics in 3 Dimensions (ASTRO 3D), Australia}
\affiliation{Center for Astrophysics $|$ Harvard \& Smithsonian, 60 Garden Street, Cambridge, MA 02138, USA}

\author[0000-0002-6620-7421]{Ralph S. Sutherland}
\affiliation{Research School of Astronomy and Astrophysics, Australian National University, Australia}
\email{peixin.zhu@cfa.harvard.edu}

\begin{abstract}

Gas metallicity, ionization parameter, and gas pressure can affect the observed ratios of specific strong emission lines within galaxies. While the theoretical strong lines diagnostics for gas metallicity, ionization parameters, and gas pressure in star-forming regions are well-established, theoretical diagnostics for active galactic nuclei (AGNs) narrow line regions are still lacking. In \citet{zhu_new_2023}, we presented a new AGN model that provides the best predictions for observations spanning the UV, optical, and infrared wavelengths. {\color{black} This paper presents a suite of theoretical diagnostics for the gas metallicity, ionization parameter, gas pressure, and the peak energy in AGN ionizing radiation field $E_{peak}$ for AGN narrow-line regions spanning the UV and optical wavelengths.} We investigate the model dependency on the ionization parameter, gas pressure, $E_{peak}$, and the nitrogen scaling relation and make recommendations on metallicity diagnostics that are most robust against these parameters. {\color{black}We test our new AGN metallicity diagnostics using optical galaxy spectra from Sloan Digital Sky Survey DR16. These tests show that the metallicities measured from different diagnostics in this paper are consistent within $\sim0.3$\,dex}. We compare consistent HII and AGN diagnostics and demonstrate that HII and AGN diagnostics should not be used interchangeably. With a wide wavelength coverage, we anticipate that these AGN diagnostics will enable new metallicity studies of galaxies dominated by AGN. 

\end{abstract}


\section{Introduction}
	\label{sec:intro}

The chemical abundances of galaxies can provide powerful constraints on theories of galaxy evolution \citep[e.g.][]{tremonti_origin_2004,rupke_galaxy_2010,ma_origin_2016,torrey_evolution_2019}. While the variation of metallicity among different types of galaxies sheds light on the galactic chemical evolution \citep[e.g.][]{kewley_host_2006,ellison_galaxy_2008,troncoso_metallicity_2014}, the distribution of chemical abundance within a single galaxy provides a complex fossil record of the star formation, stellar and AGN feedback, gas inflows, and mergers in a galaxy's history \citep[e.g.][]{kewley_metallicity_2010,torrey_metallicity_2012,ho_metallicity_2015,sharda_physics_2021}. 

Galaxy chemical abundances are traced through galaxy spectra. The emission lines and some absorption lines in galaxy spectra reflect the composition, density, and temperature of the interstellar medium in the galaxies. Using empirical and theoretical diagnostics, emission lines from HII regions have been well-studied through empirical \citep[e.g.][]{pilyugin_abundance_2003,berg_chaos_2015,2017PASP..129h2001P}, semi-empirical \citep{1979MNRAS.189...95P,maiolino_amaze_2008}, and theoretical methods \citep[e.g.][]{1991ApJ...380..140M,kewley_using_2002,nagao_metallicity_2011,grasha_metallicity_2022} to interpret the metallicity distribution in star-forming galaxies. 

{\color{black}
Using metallicity diagnostics for HII regions, a correlation between stellar mass and metallicity for star-forming galaxies was found \citep[e.g.][]{1994ApJ...420...87Z} and later confirmed and measured using local star-forming galaxies in the Sloan Digital Sky Survey (SDSS) \citep[e.g.][]{tremonti_origin_2004,kewley_metallicity_2008,mannucci_fundamental_2010,curti_mass-metallicity_2020} and high-redshift star-forming galaxies \citep{wuyts_consistent_2014,sanders_mosdef_2021}. By comparing with simulations, the mass-metallicity relation provides crucial constraints on galaxy evolution theory \citep[e.g.][]{torrey_evolution_2019,garcia_interplay_2024}.

However, most current metallicity studies are missing AGN galaxies due to the lack of consistent AGN and star-forming galaxy metallicity diagnostics. In addition to star-forming galaxies, studies on the SDSS DR16 galaxies spectra show that at least $\sim5\%$ galaxies are Seyfert galaxies and $\sim10\%$ galaxies are composite galaxies (where AGN contributes $\sim10\%-50\%$ to the emission lines) \citep{zhang_machine-learning_2019,ahumada_16th_2020,zhu_new_2023}. To obtain reliable chemical measurements for these galaxies, consistent metallicity measurements for different regions in galaxies, both HII and AGN regions, are required. }

In contrast to the metallicity measurements for HII regions, which are well-established \citep[see review by][]{kewley_understanding_2019} and tested with a large sample of star-forming galaxies in a wide range of wavelengths \citep[e.g.][]{maiolino_amaze_2008,nagao_metallicity_2011,2017PASP..129h2001P,grasha_metallicity_2022}, the metallicity measurements for the AGN regions are limited in amount and wavelength coverage. 

{\color{black}\citet{thomas_massmetallicity_2019} measure the metallicity for 7670 Seyfert 2 galaxies from SDSS DR7 and find the mass-metallicity relation for these Seyfert galaxies has a 0.09 dex systematic offset in oxygen abundance compared to the mass-metallicity derived from star-forming galaxies. Using rest-frame UV emission lines, another study on 24 high redshift ($1.2<z<4.0$) type-2 AGNs reports a positive mass-metallicity relation \citep{matsuoka_mass-metallicity_2018}.}

In the literature, metallicity measurements in the NLRs of AGNs are generally performed in three ways: the $T_e$ method, the central intersect of radial abundance gradients, and the theoretical metallicity diagnostics.

 The traditional method proposed to ``directly'' measure the gas metallicity in the interstellar medium (ISM) is the $T_e$ method \citep[e.g.][]{osterbrock_optical_1975,2006agna.book.....O,perez-montero_ionized_2017}. This method uses temperature-sensitive auroral lines such as the [O~III]$\,\lambda4363$ line to determine metallicities. This $T_e$ method was first applied to HII regions where auroral lines are observed \citep[e.g.][]{peimbert_temperature_1967,berg_carbon_2016}, and later extended to AGNs for metallicity measurement in the NLRs \citep{izotov_active_2008,dors_central_2015}. However, due to the strong temperature dependency of auroral lines, they are predominantly produced in hot gas \citep{dinerstein_abundances_1990}, which results in a bias towards the low-density and low-metallicity (12+$\log\rm(O/H)<8.7$) galaxies \citep{stasinska_biases_2005, dors_central_2015}. These auroral lines are faint and are also more difficult to observe at higher redshift, preventing the large-scale application of $T_e-$method to high$-z$ galaxies.

An alternative method proposed to estimate the metallicity in NLRs is to adopt the central intersect abundance of the radial abundance gradient derived in the outer star-forming regions of the galaxy as an approximation \citep{1992MNRAS.259..121V,1994ApJ...420...87Z,pilyugin_maximum_2007}. This approach relies on the assumption that galaxies have constant metallicity gradients from the center to the outskirts. However, as the gas-phase metallicity gradients are very sensitive to the environment and gas infall, this assumption may not be correct, especially for galaxy mergers \citep{kewley_metallicity_2010,torrey_metallicity_2012,kewley_understanding_2019}.

The most popular method to measure metallicity in NLRs, regardless of the detection of weak auroral lines, is through theoretical metallicity diagnostics. These diagnostics utilize the calibrations between the strong emission line ratios and the metallicity derived from the theoretical model of NLRs \citep[e.g.][]{storchi-bergmann_chemical_1998,groves_dusty_2004-1,feltre_nuclear_2016}. Purely based on theoretical AGN models, these theoretical metallicity calibrations can provide measurements for galaxies in a wide range of metallicity ($8.0\lesssim$12+$\log\rm(O/H)\lesssim9.3$) \citep{thomas_interrogating_2018}. Since strong emission lines are prominent in the spectra of NLRs, theoretical metallicity diagnostics are easy to apply even for galaxies at high redshift \citep[e.g.][]{trump_physical_2023,cameron_jades_2023-1}.

The most reliable and consistent metallicity measurements in NLRs are from theoretical metallicity diagnostics. A comparison performed by \citet{dors_central_2015} shows that the $T_e$ method tends to systematically underestimate the oxygen abundance of NLRs at an average level of $\sim0.8\,$ dex and produce a large fraction of sub-solar metallicity objects in the galaxy sample, while the abundances obtained by central intersect of radial metallicity gradients in \citet{2015MNRAS.450.3254P} are close to, within $\sim0.3\,$dex of the abundances derived from theoretical metallicity diagnostics \citep{1997ApJS..112..315H}. 

Nevertheless, only a few theoretical metallicity calibrations for NLRs were available in the literature, many of which strongly depend on gas ionization parameter $U$ and gas electron density $n_e$.

The first optical strong-line theoretical calibrations for NLRs are proposed by \citet{storchi-bergmann_chemical_1998}, who used the early version of CLOUDY photoionization models to study the relations between the oxygen abundances and three optical emission-line ratios [O~III]$\,\lambda\lambda4959, 5007$/H$\beta$, [N~II]$\,\lambda\lambda6548, 6584$/H$\alpha$, and [O~II]$\,\lambda3727$/[O~III]$\,\lambda\lambda4959, 5007$. They obtained two AGN metallicity calibrations. One involves the linear combinations of [N~II]/H$\alpha$ and [O~III]/H$\beta$ and is relatively insensitive to reddening. The other uses the decimal logarithms of [N~II]/H$\alpha$ and [O~II]/[O~III] and is less sensitive to the ionization parameter. \citet{castro_new_2017} using CLOUDY 13 code, proposed [N~II]$\,\lambda6584$/[O~II]$\,\lambda3727$ as a mostly ionization parameter independent metallicity diagnostic for NLRs. Also, using CLOUDY 17, \citet{carvalho_chemical_2020} produced a semi-empirical metallicity calibration using the optical emission-line ratio [N~II]$\,\lambda6584$/H$\alpha$, which requires prior knowledge of $U$, $n_e$, and the spectral index of the AGNs $\alpha_{ox}$. To overcome the degeneracy between metallicity and ionization parameters, attempts have also been made to estimate metallicity and ionization parameter simultaneously by applying a Bayesian approach based on AGN photoionization models \citep{thomas_interrogating_2018, perez-montero_bayesian-like_2019}.

The first UV metallicity calibration was explored by \citet{nagao_gas_2006} to measure the metallicity in high$-z$ galaxies. Using the Cloudy 94 code, they found that the strong emission-line ratio C~IV$\,\lambda1549$/He~II$\,\lambda1640$ can be used as a metallicity diagnostic and that C~III]$\,\lambda1909$/C~IV$\,\lambda1549$/ can be used as an ionization parameter sensitive diagnostic. However, this metallicity calibration is non-monotonic and depends strongly on the AGN ionizing spectrum in the NLR model. \citet{2014MNRAS.443.1291D} proposed (C~III]$\,\lambda1909$+C~IV$\,\lambda1549$)/He~II$\,\lambda1640$ (C43) as a metallicity indicator for NLRs with moderate dependence on ionization parameter. To mitigate the effect of ionization parameter, \citet{dors_semi-empirical_2019} combined the emission-line ratio C43 with two ionization parameter sensitive emission-line ratios N~V$\,\lambda1549$/He~II and C~III]/C~IV separately to derive two semi-empirical metallicity calibrations for NLRs.

In addition to being limited in quantity and wavelength coverage, these AGN metallicity diagnostics show discrepancies when applied to the same AGN spectra. Using the optical spectra of 463 confirmed Seyfert 2 AGNs, \citet{dors_chemical_2020-1} compare the metallicities measured from two theoretical diagnostics, one using the combination of [O~III]/H$\beta$ and [N~II]/H$\alpha$ from \citep{storchi-bergmann_chemical_1998} and the other using [N~II]/[O~II] from \citep{castro_new_2017}. They find the measurements from these two AGN metallicity diagnostics differ by up to $\sim0.5$ dex for both high-metallicity (12+$\log\rm(O/H)>8.9$) and low-metallicity (12+$\log\rm(O/H)<8.5$) Seyfert galaxies. Even with the same emission line ratio [N~II]/[O~II], the metallicity diagnostic derived by \citet{thomas_interrogating_2018} gives $\sim0.4$ dex higher metallicity measurements than the metallicity diagnostics derived by \citet{castro_new_2017}.

The discrepancy among the AGN metallicity diagnostics is due to the differences among the AGN models. A typical AGN model contains an AGN ionization spectrum, descriptions of gas abundance, gas density structure, and dust properties, and a photoionization code to perform numerical calculations. The selection of model settings in each of these factors affects the final AGN metallicity diagnostics \citep[e.g.][]{groves_dusty_2004-1,feltre_nuclear_2016}. Despite the known differences, it was unclear to what extent NLR metallicity diagnostics are reliable and which input parameters should be used for a large sample of AGNs until recently. 

Recently, in \citet{zhu_new_2023}, we resolved the discrepancies by investigating how the selection of each input parameter affects the predictions of AGN theoretical diagnostics. Based on comparison with observations, we recommended a new AGN model that can consistently predict observed UV, optical, and IR emission lines in NLRs.

In this paper, we aim to overcome the shortage of methods to measure metallicity in the AGN regions by providing a set of consistent AGN metallicity diagnostics across a wide coverage of wavelengths. We perform a comprehensive search for AGN metallicity diagnostics in UV and optical wavelengths with the new AGN models from \citet{zhu_new_2023}, aiming to enrich the number of metallicity diagnostics across a broad wavelength coverage in the NLR regions of AGN.  Because our new diagnostics use the MAPPINGS photoionization models and consistent abundance sets, our new AGN NLR diagnostics may be used in conjunction with the metallicity diagnostics for star-forming galaxies presented in \citet{kewley_understanding_2019}. The broad wavelength coverage of our AGN theoretical diagnostics will provide a consistent means of interpreting the spectra of galaxies from the nearby universe to the high-redshift universe, enabling the studies of chemical evolution across different redshift.

This paper is structured as follows. We first briefly introduce the AGN models in Section 2. Sections 3 to 6 present AGN theoretical diagnostics for metallicity, ionization parameter, gas pressure, and AGN radiation field, respectively. Diagnostic testing is given in Section 7. {\color{black}The bias of metallicity measurement in HII and AGN mixed galaxies is discussed in Section 8, followed by conclusion in Section 9.}

\section{Theoretical Models}
	\label{sec:model}
	
\subsection{Radiation Field}

We adopt the physically based radiation model OXAF \citep{thomas_physically_2016} to generate AGN ionizing radiation fields for our photoionization models. As a simple version of a more complicated radiation model OPTXAGNF \citep{done_intrinsic_2012,jin_combined_2012-1}, OXAF characterizes the continuum emission radiated from a thin accretion disk and a Comptonizing corona surrounding a central rotating black hole with three parameters: the peak energy $E_{\rm peak}$ of the accretion disk emission (the big blue bump 'BBB' emission), the photon index of the inverse Compton scattered power-law tail $\Gamma$, and the flux ratio of the non-thermal tail to the total flux $p_{\rm NT}$. 

Among the three parameters, \citet{thomas_physically_2016} found that $E_{\rm peak}$ plays the most significant role in influencing the emission lines predicted by the models. In addition, $E_{\rm peak}$ is a function of the central black hole mass, AGN luminosity, and the corona radius, as shown in Function (4) in \citet{thomas_physically_2016}.

The soft X-ray excess is not included in the OXAF model. Because the soft X-ray excess photons with energy $\sim0.2-2\,$keV have less effect on the ionization state of the nebula than the photon with energy $\sim0.01-0.2\,$keV, the intermediate soft X-ray component is less important in the prediction of diagnostic optical emission-line ratios \citep{dopita_probing_2014,thomas_physically_2016}. Although dominant soft X-ray excesses do affect the abundances of high-ionization potential species such as [Fe~{\sc xiv}]361 eV and [Fe~{\sc x}]234 eV, a comprehensive discussion of this effect falls beyond the scope of this paper.

Therefore, here we use the OXAF model to generate four AGN radiation fields with fixed $\Gamma=2.0$ and $p_{\text{NT}}=0.15$ and varying $\log E_{\text{peak}}/(\text{keV})=-2.0,-1.5,-1.25,-1.0$ to represent AGN radiation fields within the range of $10^6\lesssim M_{\rm BH}/M_{\odot}\lesssim10^9$ and $0.05\lesssim L/L_{\rm Edd}\lesssim0.3$ \citep{novikov_astrophysics_1973,done_intrinsic_2012,thomas_physically_2016}.

\subsection{Photoionization Models}

We use the MAPPINGS version 5.20 photoionization code \citep{sutherland_mappings_2018} to model the interstellar medium in the NLR surrounding the AGNs, noting the discrepancy between the diagnostic optical emission-line ratios calculated by CLOUDY 17.0 \citep{ferland_2017_2017} and MAPPINGS V is generally within 0.1$\,$dex \citet{zhu_new_2023}, with the exception for [S~{\sc ii}]$\,\lambda\,\lambda$6717,31/H$\alpha$ (which can go up to $\sim0.3\,$dex due to different atomic data adopted for SII). The latest available atomic data from CHIANTI version 10 \citep{del_zanna_chiantiatomic_2021} is included in MAPPINGS V. 

To characterize the relative abundance of individual elements in AGN models, we adopt the abundance set from \citet{2017MNRAS.466.4403N}. An abundance set consists of a standard abundance reference and abundance scaling relations describing how the relative abundance for the other 29 elements to hydrogen abundance changes with metallicity. Instead of using the solar abundance as the standard abundance reference, \citet{2017MNRAS.466.4403N} use the standard local B-stars abundance from \citet{nieva_present-day_2012}. This 	`Cosmic Abundance Standard' proposed by \citet{nieva_present-day_2012} is based on the average abundance of 29 local B stars in the Milky Way, as the B-star photospheric abundances represent the bulk abundances of the nebular region in which they recently formed. In addition to the eight elements He, C, N, O, Ne, Mg, Si, and Fe provided by the `Cosmic Abundance Standard', \citet{2017MNRAS.466.4403N} supplement the abundance of additional 22 elements with recent solar and meteoric abundances \citep{lodders_abundances_2009,grevesse_elemental_2015,scott_elemental_2015-1,scott_elemental_2015}. 

In \citet{2017MNRAS.466.4403N}, the abundance scaling relations for most elements are measured to be linear, except for helium, carbon, and nitrogen. For helium, the stellar yield in addition to the primordial abundance from \citet{pagel_primordial_1992} is adopted. For carbon and nitrogen, primary and secondary nucleosynthetic components are used to fit the stellar abundances observed from nearby HII regions \citep[e.g.][]{spite_first_2005, nieva_present-day_2012}. The non-linear scaling relations of carbon used by \citet{2017MNRAS.466.4403N} are consistent with other literature \citep[e.g.][]{berg_carbon_2016,berg_chemical_2019}. However, disagreement about the nitrogen scaling relation N/O$-$O/H remains, mainly about the oxygen abundance where the secondary nucleosynthesis contributed by intermediate-mass stars and AGB stars starts to dominate over the primary nucleosynthesis originating from core-collapse supernovae \citep{dors_new_2017,perez-montero_bayesian-like_2019,carvalho_chemical_2020}. 

The two most frequently used N/O$-$O/H relations assume the transition oxygen abundance of 12+$\log(\rm O/H)\approx8.0$ \citep[`NHhigh';][]{dors_new_2017,carvalho_chemical_2020} and 12+$\log(\rm O/H)\approx8.2$ \citep[`NHlow'][]{groves_dusty_2004-1,gutkin_modelling_2016,2017MNRAS.466.4403N}, respectively. The slope of the secondary component in the `NHhigh' relation ($\alpha=1.29$) is larger than in the `NHlow' relation ($\alpha=1$), as shown in the equation (1) for `NHhigh' and the equation (2) for `NHlow'. Therefore, at the high metallicity end (12+$\log(\rm O/H)\approx9.0$), the N/O ratio derived from the `NHhigh' N/O$-$O/H relation can be $\sim0.4$\,dex higher than the N/O ratio derived from the `NHlow' N/O$-$O/H relation. This effect will pass on to the predictions of the emission line ratios that contain nitrogen emission lines, including {[N~{\sc ii}]}/H$\alpha$ and {[N~{\sc ii}]}/{[O~{\sc ii}]} which are frequently used as metallicity diagnostics. 

\begin{small}
\begin{equation}\nonumber
\text{`NHlow'}:\log(\rm N/O)=\log(10^{-1.732}+10^{[\log(O/H)+2.19]})
\end{equation}
\begin{equation}\nonumber
\text{`NHhigh'}:\log(\rm N/O)=1.29\times(12+\log(O/H))-11.84
\end{equation}
\end{small}

The physics behind the non-uniform transiting oxygen abundance is the star formation history of galaxies. {\color{black} The enrichment of nitrogen consists of a primary source from core-collapse supernovae in the native gas cloud \citep{van_zee_abundances_1998} and a secondary source from the delayed nucleosynthesis from intermediate-mass stars in their giant phase \citep{renzini_advanced_1981}. Primary nucleosynthesis dominates at low metallicity and leads to a constant N/O ratio because massive stars that lead to core-collapse supernovae have shorter main-sequence lifetimes than intermediate-mass stars. As intermediate-mass stars evolve off the main sequence, the secondary nucleosynthesis dominates and leads to a rapidly growing N/O ratio. The timescale difference between primary and secondary nucleosynthesis results in a transition metallicity in the N/O$-$O/H scaling relations where the secondary nucleosynthesis starts to dominate over the primary nucleosynthesis \citep{karakas_dawes_2014,cameron_nitrogen_2023}

 The transition metallicity can be intrinsically different in galaxies, depending on the individual star formation history of a galaxy \citep{nomoto_nucleosynthesis_2013,kobayashi_origin_2020}.} For galaxies that formed a significant number of Wolf-Rayet stars in their star formation history, the secondary nitrogen enrichment will dominate early when galaxies are still relatively metal-poor, resulting in a lower transition metallicity \citep{berg_re-examining_2011,masters_physical_2014}. 

Given the intrinsic variation in the N/O$-$O/H relation among different types of galaxies, we decided to build two versions of AGN models using the `NHlow' and `NHhigh' relations, respectively. As most of the stellar and nebular nitrogen abundance observations lie between the `NHlow' and `NHhigh' relations on the $\log(\rm N/O)-\log(O/H)$ diagram \citep{dors_new_2017,2017ApJ...840...44B,perez-montero_bayesian-like_2019,berg_chemical_2019}, the predictions from the two versions of AGN models can provide an upper limit and a lower limit for reference on the intensity of nitrogen emission line observed in AGNs. For readers who wish to use the nitrogen-based diagnostics introduced in the following sections, we caveat that the N/O$-$O/H relation in the AGN model should be carefully determined in advance based on the class of objects to which the diagnostics are applied.

We include dust in our AGN models and adopt the depletion factors derived from \citet{jenkins_unified_2009}, assuming iron to be 97.8\% depleted. The dust destruction process is not considered in AGN models, as the charge of dust grains is too low ($q<5V$) to trigger the dust destruction process (requires $q\approx100V$, \citet{dopita_astrophysics_2003}).

For the density structure, we assume pressure equilibrium and allow the density to vary within the NLR. The pressure equilibrium ensures a constant total pressure across the NLR, which is the sum of radiative pressure introduced by dust grains and gas pressure. Constant pressure (isobaric) is a more realistic assumption than constant density, as the constant density assumption in the NLR is violated by observations \citep{binette_excitation_1996,fernandez-ontiveros_far-infrared_2016}.

We use plane-parallel geometry in our AGN models. To characterize the total radiative flux entering the photoionization cloud, we adopt the dimensionless ionization parameter at the inner edge of the cloud $U=S_{\star}/(n_0c)$ as the ionization parameter used in our AGN models, where the $S_{\star}$ represents the flux of ionizing photons and the initial value of the gas density $n_0$ is set by the initial gas pressure $P_0$ at the inner edge of the cloud. This definition is chosen to facilitate comparison between our diagnostics and other diagnostics published in the literature \citep{groves_dusty_2004-1,castro_new_2017,dors_semi-empirical_2019,carvalho_chemical_2020}.

Detailed photoionization, excitation, and recombination are calculated at a step size of 0.02 throughout the NLR. The calculation stops when the hydrogen becomes 99\% neutral.

To develop the metallicity diagnostics for the NLR in AGNs and study how these diagnostics depend on ionization parameters and gas pressure, we calculate three-dimensional grids of AGN models with each of the fixed AGN radiation fields at $\log E_{\text{peak}}/(\text{keV})$=(-2.0, -1.5, -1.25, -1.0). The three-dimensional grids of AGN models are calculated by holding the gas pressure fixed at $\log{P}$=(6.2, 7.0, 7.8) and the ionization parameter fixed at each of $\log(U)$=(-3.8, -3.4,-3.0,-2.6,-2.2,-1.8,-1.4,-1.0) while varying the metallicity in the range of 12+$\log(\rm O/H)$=(7.297, 7.695, 7.996, 8.281, 8.434, 8.542, 8.695, 8.804, 8.888, 8.957, 9.015, 9.093, 9.164, 9.261, 9.34, 9.407). We calculate two sets of AGN models with `NHlow' and `NHhigh' scaling relations to account for the intrinsic variation in nitrogen scaling relations.

%Section 2: draft finished on Sep 17.

\section{metallicity diagnostics}
	\label{sec:metal}
%check !! for completeness before submit
	
This section presents metallicity diagnostics for the NLR in AGNs in UV and optical wavelengths, respectively. Similar to the situation in HII regions, many AGN metallicity diagnostics have small to moderate dependence on the ionization parameter ($\log(U)$), gas pressure ($P/k$), and the peak energy ($E_{\rm peak}$) of the AGN radiation field. For convenience in usage, we discuss their valid ranges in gas phase metallicity and their parameter degeneracy. We also perform numerical fitting for the recommended metallicity diagnostics to facilitate easy application. {\color{black} All emission line ratios discussed in this section are in log scale.}

\subsection{UV Metallicity Diagnostics}
\begin{figure*}[htb]
\epsscale{1.0}
\plotone{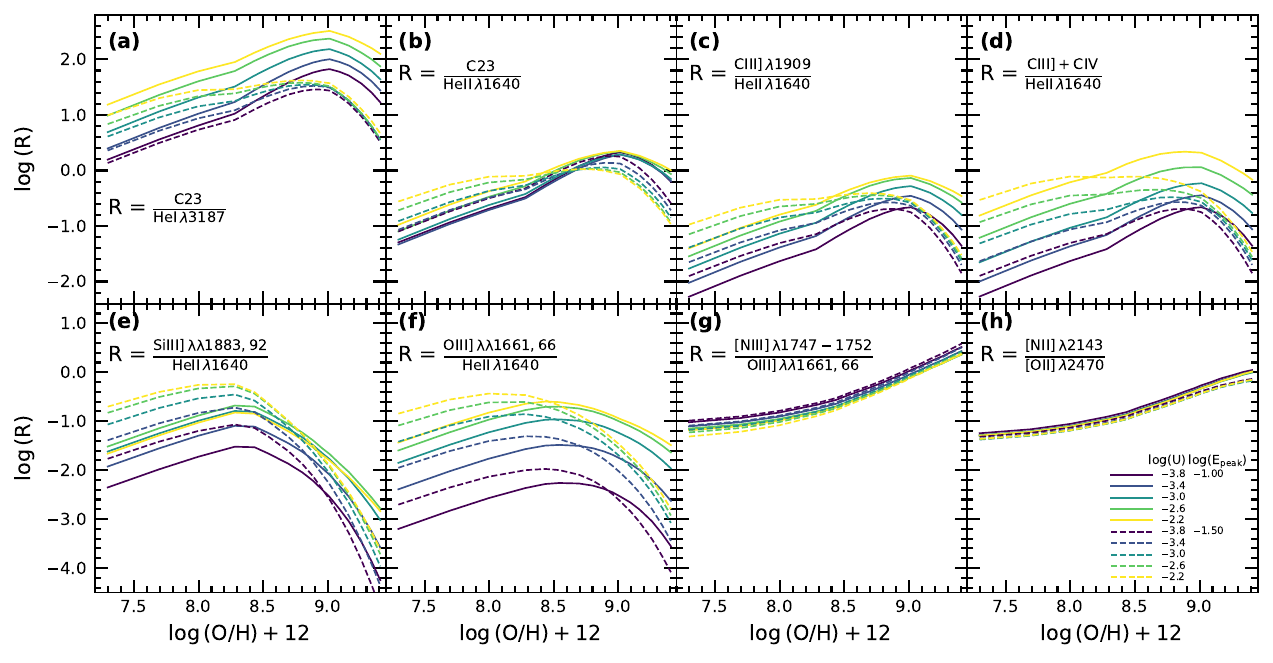}
\caption{UV metallicity diagnostics for the AGN narrow line regions as predicted by AGN models with $\log(P/k)=7.0$, `NHlow' nitrogen scaling relation, and varying ionization parameters ($\log(U)=-3.8,-3.4,-3.0,-2.6,-2.2$ correspond to colors from dark to bright) and varying peak energy in the AGN radiation field ($\log(E_{\rm peak}/\rm keV)=-1.0$ in solid lines and $\log(E_{\rm peak}/\rm keV)=-1.5$ in dash lines). 
\label{fig:1}}
\end{figure*}

\begin{figure*}[htb]
\epsscale{1.0}
\plotone{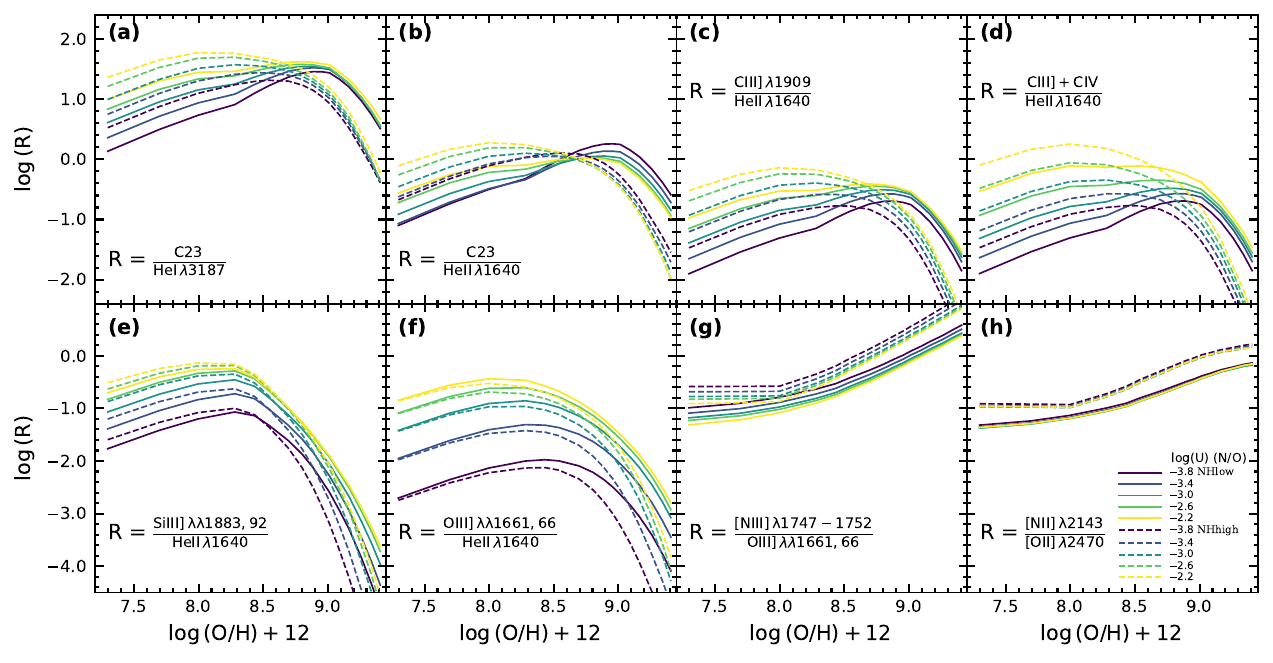}
\caption{{\color{black}The effect of nitrogen scaling relation on UV metallicity diagnostics for the AGN narrow line regions as predicted by AGN models with $\log(P/k)=7.0$ and $\log(E_{\rm peak}/\rm keV)=-1.5$.} As in Figure~\ref{fig:1}, but with line styles representing the nitrogen scaling relation used in the AGN models. (`NHlow' in solid lines and `NHhigh' in dash lines) 
\label{fig:B1}}
\end{figure*}

The most frequently observed UV emission lines in NLRs are C~IV$\,\lambda1549$, He~II$\,\lambda1640$, O~III]$\,\lambda\lambda1661,66$, [C~III]$\,\lambda1907$, and C~III]$\,\lambda1909$. In addition to these emission lines, we also investigate metallicity diagnostics that evolve other emission lines that are less frequently detected in previous UV observations but are detectable with JWST: Si~III]$\,\lambda\lambda1883,92$, N~III] complex (the sum of $\,\lambda1747$, $\,\lambda1748$, $\,\lambda1749$, $\,\lambda1752$, and $\,\lambda1754$), [N~II]$\,\lambda2143$, [C~II]$\,\lambda2325$ blend (the sum of $\,\lambda2323.5$, $\,\lambda2324.69$, $\,\lambda2325.4$, $\,\lambda2326.93$, and $\,\lambda2328.12$), [O~II]$\,\lambda2470$, and He~I$\,\lambda3187$.

Figure~\ref{fig:1} shows the most promising UV metallicity diagnostics in NLRs derived from AGN models with $\log(P/k)=7.0$ and `NHlow' nitrogen scaling relation, where colors correspond to ionization parameters and line styles reflect the peak energy ($E_{\rm peak}$)in the AGN radiation field. Table~\ref{tab:1} gives the coefficient fits for these diagnostics regarding the ionization parameter for each $E_{\rm peak}$ value. {\color{black} The coefficient fits for the same set of diagnostics but derived from AGN models with `NHhigh' nitrogen scaling relation can be found in Table~\ref{tab:C1}.}

We use He~II$\,\lambda1640$ and He~I$\,\lambda3187$ as the baselines for UV metallicity diagnostics, as there are no useful hydrogen lines at UV wavelengths. Unlike star-forming regions, the He~I$\,\lambda3187$ in NLRs are more than 10x dimmer than He~II$\,\lambda1640$. The He~II$\,\lambda1640$ line, with a high ionization potential (54.4 eV), is very sensitive to the EUV radiation field and ionization parameter. As AGN generates a much harder spectrum than star-forming region, the He~II$\,\lambda1640$ in NLRs is much brighter than in HII regions. \citep{shirazi_strongly_2012,tozzi_unveiling_2023} It is important to keep in mind that shocks can also generate hard EUV radiation and contribute to the He~II$\,\lambda1640$ lines \citep{sutherland_effects_2017,dopita_effects_2017}.

UV wavelengths are rich in carbon emission lines. The $C_{23}$/He~II ratio (Figure~\ref{fig:1}b), {\color{black}where $C_{23}$ is the sum of [C~II]$\,\lambda2325$ blend, [C~III]$\,\lambda1907$, and C~III]$\,\lambda1909$}, is a useful metallicity diagnostic for AGN with $\log (E_{\rm peak})>-1.5$, with only moderate dependence on ionization parameter ($C_{23}$/He~II varies up to $\sim0.4\,$dex for ionization parameters $-3.8<\log(U)<-2.2$). Both $C_{23}$/He~I and C~III]/He~II ratios (Figure~\ref{fig:1}a,c) have a larger dependence on the ionization parameter ($C_{23}$/He~I and C~III]/He~II varies up to $\sim1$\,dex for ionization parameters $-3.8<\log(U)<-2.2$). With the inclusion of C~IV$\,\lambda1549$, the $C_{34}$/He~II ratio (Figure~\ref{fig:1}d) has an increased dependence on the ionization parameter ($C_{34}$/He~II varies up to $\sim1.2\,$dex for ionization parameters $-3.8<\log(U)<-2.2$), suggesting $C_{23}$/He~II and C~III]/He~II (if [C~II] is not available) are more robust metallicity diagnostics than $C_{34}$/He~II if the ionization parameter cannot be determined. 

{\color{black}As a major constituent of grains, the abundance of carbon in the gas phase is sensitive to the dust depletion factor used in the AGN model. The best estimations so far on dust depletion factor from \citet{jenkins_unified_2009,jenkins_depletions_2014} are used in our AGN model. Nevertheless, we caveat that these carbon diagnostics might be affected if the carbon depletion factor is updated in the future.}
 
\begin{deluxetable*}{c|c|c|c|c|c|c|c|c}[hbt]
\tabletypesize{\scriptsize}
\tablewidth{0pc}
\tablecaption{Cubic Surface Fits for UV Metallicity Diagnostics for AGN Narrow Line Regions with $\log(P/k)=7.0$ and `NHlow' Scaling Relation\label{tab:1}}
\tablenum{1}
\tablehead{\multicolumn{9}{c}{$F=a_1X+a_2Y+a_3Z+a_4X^2+a_5XY+a_6XZ+a_7Y^2+a_8YZ+a_9Z^2+a_{10}X^3+a_{11}X^2Y$}\\
\multicolumn{9}{c}{$+a_{12}X^2Z+a_{13}XY^2+a_{14}XYZ+a_{15}XZ^2+a_{16}Y^3+a_{17}Y^2Z+a_{18}YZ^2+a_{19}Z^3+a_{20}$,}\\ 
\multicolumn{9}{c}{where $X=\log(U)$, $Y=\log(E_{\rm peak}/\rm keV)$, $Z=\log(R)$, $F=\log(\rm O/H)+12$,$-3.8\leq X\leq-1.0$, $-2.0 \leq Y\leq-1.0$}\\
\hline
\colhead{$\log$(R)} & \colhead{$\rm \frac{C_{23}}{HeI\,\lambda3187}$\tablenotemark{\scriptsize a}} & \colhead{$\rm \frac{C_{23}}{HeII\,\lambda1640}$} & \colhead{$\rm \frac{CIII]\,\lambda1909}{HeII\,\lambda1640}$} & \colhead{$\rm \frac{C_{43}}{HeII\,\lambda1640}$} & \colhead{$\rm \frac{SiIII]\,\lambda\lambda1883,92}{HeII\,\lambda1640}$} & \colhead{$\rm \frac{OIII]\,\lambda\lambda1661,66}{HeII\,\lambda1640}$} & \colhead{$\rm \frac{[NIII]\,\lambda1747-1752}{OIII]\,\lambda\lambda1661,66}$} & \colhead{$\rm \frac{[NII]\,\lambda2143}{[OII]\,\lambda2470}$}}
\startdata
$F_{\rm min}$ & 8.96 & 8.96 & 8.96 & 8.96 & 8.54 & 8.96 & 7.70 & 7.70 \\
$F_{\rm max}$ & 9.41 & 9.41 & 9.41 & 9.41 & 9.41 & 9.41 & 9.41 & 9.41 \\
\hline
$a_1$ & 0.0078 & 1.3428 & 0.8415 & 0.6448 & -1.1518 & 0.5104 & 0.0443 & 0.5171 \\
$a_2$ & 4.1113 & 12.7000 & 11.8426 & 8.4525 & -3.1689 & 7.4426 & 0.0769 & -5.0600 \\
$a_3$ & 2.0899 & 0.0378 & -1.8191 & 0.1206 & -1.7150 & -2.1638 & 1.1622 & 0.7214 \\
$a_4$ & 0.0417 & 0.0896 & 0.0371 & -0.0673 & -0.0728 & 0.1629 & 0.1075 & 0.2388 \\
$a_5$ & -0.2260 & 1.3007 & 0.9975 & 0.5447 & -0.7502 & 0.6588 & -0.1899 & 0.0448 \\
$a_6$ & 0.3611 & -0.0958 & -0.2518 & 0.2212 & -0.2295 & -0.3777 & -0.0508 & -0.1628 \\
$a_7$ & 3.1339 & 6.1916 & 6.1104 & 3.9806 & -1.0751 & 4.2366 & 0.4750 & -3.4376 \\
$a_8$ & 1.5989 & 1.9147 & 0.0125 & 0.7996 & -0.7050 & -0.8011 & 0.6041 & -0.4447 \\
$a_9$ & -0.1073 & -1.3376 & -0.7544 & -0.4048 & -0.1563 & -0.2218 & 0.5611 & 1.0541 \\
$a_{10}$ & 0.0133 & 0.0144 & 0.0315 & 0.0177 & 0.0334 & 0.0701 & 0.0159 & 0.0338 \\
$a_{11}$ & -0.0850 & -0.0495 & -0.0981 & -0.1232 & -0.1178 & -0.1142 & 0.0132 & -0.0001 \\
$a_{12}$ & -0.0234 & 0.0058 & -0.0512 & -0.0316 & -0.0360 & -0.0535 & 0.0005 & -0.0052 \\
$a_{13}$ & 0.0990 & 0.4602 & 0.4274 & 0.2690 & -0.0437 & 0.4016 & -0.0607 & 0.0455 \\
$a_{14}$ & 0.2320 & 0.0226 & 0.0378 & 0.1805 & -0.0249 & -0.0688 & -0.0150 & -0.1405 \\
$a_{15}$ & -0.0373 & -0.0757 & -0.0161 & 0.0124 & 0.0030 & -0.0002 & 0.1875 & 0.0919 \\
$a_{16}$ & 0.6967 & 0.9054 & 0.9372 & 0.5785 & -0.1938 & 0.6862 & 0.1888 & -0.9180 \\
$a_{17}$ & 0.1830 & 0.7355 & 0.1746 & 0.1888 & -0.0890 & -0.1420 & 0.2015 & 0.5105 \\
$a_{18}$ & 0.0140 & -0.4307 & -0.2283 & -0.1179 & -0.0421 & -0.0112 & 0.1065 & -0.9339 \\
$a_{19}$ & -0.0639 & -0.1227 & -0.0482 & -0.0364 & -0.0072 & -0.0182 & 0.1491 & 1.2253 \\
$a_{20}$ & 11.5151 & 17.4070 & 15.8565 & 15.0513 & 4.7555 & 12.0347 & 8.9746 & 7.1936 \\
\hline
RMS err(\%) & 2.20\% & 2.59\% & 2.64\% & 2.81\% & 2.16\% & 2.21\% & 2.56\% & 3.22\% \\
\enddata
\tablenotetext{\tiny a}{$C_{23}$=[C~II]$\,\lambda2325$ blend+[C~III]$\,\lambda1907$+C~III]$\,\lambda1909$}
\end{deluxetable*}
\begin{deluxetable*}{c|c|c|c|c|c|c|c|c}[hbt]
\tabletypesize{\scriptsize}
\tablewidth{0pc}
\tablecaption{Cubic Surface Fits for UV Metallicity Diagnostics for AGN Narrow Line Regions with $\log(P/k)=7.0$ and `NHhigh' Scaling Relation\label{tab:C1}}
\tablenum{2}
\tablehead{\multicolumn{9}{c}{$F=a_1X+a_2Y+a_3Z+a_4X^2+a_5XY+a_6XZ+a_7Y^2+a_8YZ+a_9Z^2+a_{10}X^3+a_{11}X^2Y$}\\
\multicolumn{9}{c}{$+a_{12}X^2Z+a_{13}XY^2+a_{14}XYZ+a_{15}XZ^2+a_{16}Y^3+a_{17}Y^2Z+a_{18}YZ^2+a_{19}Z^3+a_{20}$,}\\ 
\multicolumn{9}{c}{where $X=\log(U)$, $Y=\log(E_{\rm peak}/\rm keV)$, $Z=\log(R)$, $F=\log(\rm O/H)+12$,$-3.8\leq X\leq-1.0$, $-2.0 \leq Y\leq-1.0$}\\
\hline
\colhead{$\log$(R)} & \colhead{$\rm \frac{C_{23}}{HeI\,\lambda3187}$\tablenotemark{\scriptsize a}} & \colhead{$\rm \frac{C_{23}}{HeII\,\lambda1640}$} & \colhead{$\rm \frac{CIII]\,\lambda1909}{HeII\,\lambda1640}$} & \colhead{$\rm \frac{C_{43}}{HeII\,\lambda1640}$} & \colhead{$\rm \frac{SiIII]\,\lambda\lambda1883,92}{HeII\,\lambda1640}$} & \colhead{$\rm \frac{OIII]\,\lambda\lambda1661,66}{HeII\,\lambda1640}$} & \colhead{$\rm \frac{[NIII]\,\lambda1747-1752}{OIII]\,\lambda\lambda1661,66}$} & \colhead{$\rm \frac{[NII]\,\lambda2143}{[OII]\,\lambda2470}$}}
\startdata
$F_{\rm min}$ & 8.96 & 8.89 & 8.89 & 8.96 & 8.54 & 8.96 & 7.70 & 7.70 \\
$F_{\rm max}$ & 9.41 & 9.41 & 9.41 & 9.41 & 9.41 & 9.41 & 9.41 & 9.41 \\
\hline
$a_1$ & 0.8362 & 0.8918 & 0.6847 & 0.5680 & -1.0584 & 0.4350 & 0.0348 & 0.1490 \\
$a_2$ & 3.9997 & 7.8230 & 7.3373 & 5.4144 & -3.0121 & 4.6715 & -0.6968 & -4.2028 \\
$a_3$ & -0.1978 & -0.5295 & -1.2488 & -0.0499 & -1.6648 & -1.3298 & 0.5678 & 0.0004 \\
$a_4$ & 0.0547 & -0.0387 & -0.0369 & -0.1000 & -0.0681 & 0.0689 & 0.0868 & 0.1559 \\
$a_5$ & 0.5692 & 1.1037 & 1.0054 & 0.6119 & -0.6396 & 0.6240 & -0.1988 & -0.2307 \\
$a_6$ & 0.0585 & -0.1425 & -0.2030 & 0.0685 & -0.2621 & -0.2137 & -0.1772 & -0.2013 \\
$a_7$ & 1.5856 & 3.2531 & 3.2662 & 2.0868 & -1.0868 & 2.2405 & -0.0449 & -2.7730 \\
$a_8$ & -0.1981 & 0.9739 & 0.1244 & 0.5111 & -0.7006 & -0.3021 & 0.1691 & -0.4476 \\
$a_9$ & 0.1926 & -0.6808 & -0.4072 & -0.1862 & -0.1531 & -0.1802 & 0.2019 & -0.2162 \\
$a_{10}$ & 0.0067 & -0.0023 & 0.0154 & 0.0081 & 0.0275 & 0.0475 & 0.0144 & 0.0266 \\
$a_{11}$ & -0.0327 & -0.0365 & -0.0637 & -0.0856 & -0.0965 & -0.0914 & 0.0032 & -0.0244 \\
$a_{12}$ & -0.0276 & -0.0214 & -0.0389 & -0.0317 & -0.0296 & -0.0348 & -0.0129 & -0.0042 \\
$a_{13}$ & 0.2083 & 0.3663 & 0.3829 & 0.2575 & -0.0260 & 0.3364 & -0.0535 & -0.0085 \\
$a_{14}$ & 0.0991 & 0.0481 & 0.0145 & 0.0675 & -0.0477 & -0.0299 & -0.0199 & -0.1573 \\
$a_{15}$ & -0.0136 & -0.0470 & -0.0097 & 0.0213 & -0.0032 & -0.0006 & 0.0862 & 0.0542 \\
$a_{16}$ & 0.2121 & 0.3453 & 0.3905 & 0.2106 & -0.1997 & 0.2711 & 0.0738 & -0.7176 \\
$a_{17}$ & -0.1549 & 0.3803 & 0.1014 & 0.1048 & -0.0897 & -0.0332 & 0.0722 & 0.3979 \\
$a_{18}$ & 0.1536 & -0.1500 & -0.0483 & 0.0024 & -0.0356 & -0.0129 & 0.0645 & -0.9448 \\
$a_{19}$ & -0.0555 & -0.0711 & -0.0440 & -0.0369 & -0.0068 & -0.0136 & 0.0587 & 0.7251 \\
$a_{20}$ & 13.0812 & 14.3284 & 13.3460 & 13.2275 & 5.0648 & 11.0864 & 8.3160 & 6.9964 \\
\hline
RMS err(\%) & 3.24\% & 3.32\% & 3.56\% & 3.81\% & 2.13\% & 2.75\% & 1.83\% & 4.20\% \\
%R2_1 & 0.9447 & 0.9539 & 0.9468 & 0.9392 & 0.9931 & 0.9683 & 0.9979 & 0.9891 \\
%R2_2 & 0.7648 & 0.7852 & 0.7693 & 0.7535 & 0.9168 & 0.8219 & 0.9546 & 0.8956 \\
%score & 0.9447 & 0.9539 & 0.9468 & 0.9392 & 0.9931 & 0.9683 & 0.9979 & 0.9891 \\
\enddata
\tablenotetext{\tiny a}{$C_{23}$=[C~II]$\,\lambda2325$ blend+[C~III]$\,\lambda1907$+C~III]$\,\lambda1909$}
\end{deluxetable*}
In the $12+\log(\rm O/H)>9.0$ regime, all carbon metallicity diagnostics show a large dependence on the $E_{\rm peak}$ (the line ratio varies up to $\sim1$\,dex for $-1.5<E_{\rm peak}<-1.0$) and moderate dependence on the gas pressure (the line ratio varies up to $\sim0.6$\,dex for gas pressure $6.2<\log(P/k)<7.8$), requiring the prior knowledge of $E_{\rm peak}$ and gas pressure to estimate metallicity with the carbon lines. 

{\color{black}These carbon line ratios, although sensitive to metallicity, are double-valued with metallicity. This is a combination effect of both gas metallicity and temperature. At low metallicity ($12+\log(\rm O/H)\lesssim8.8$), the nebula has a high temperature, and thus the ionized carbon mainly scales with carbon abundance. However, at high metallicity ($12+\log(\rm O/H)\gtrsim9.0$), carbon emission lines are weakened as the collisional excitation of the carbon ions is reduced by the low temperatures in high metallicity nebulae. Because the nebular temperature is closely related to the nitrogen abundance, these carbon line ratios are also sensitive to the nitrogen scaling relations used in the AGN models, as shown in Figure~\ref{fig:B1}.} Additional line ratios are required to determine which branch the carbon metallicity diagnostics lie on. 

The Si~III]/He~II ratio (Figure~\ref{fig:1}e) is also sensitive to metallicity, especially in metal-rich regions ($12+\log(\rm O/H)>8.5$). The silicon lines are dependent on the ionization parameter (Si~III]/He~II varies up to $\sim1$\,dex for ionization parameters $-3.8<\log(U)<-2.2$), the $E_{\rm peak}$ of the AGN radiation field (Si~III]/He~II varies up to $\sim0.4\,$dex for $-1.5<E_{\rm peak}<-1.0$), and the gas pressure (Si~III]/He~II varies up to $\sim0.4\,$dex for gas pressure $6.2<\log(P/k)<7.8$). The silicon lines are affected by the gas phase abundance of silicon, which can be heavily depleted into dust grains. {\color{black}The current depletion factor used for silicon in our AGN model is -0.7\,dex at $12+\log(\rm O/H)=8.76$ \citep{jenkins_unified_2009,jenkins_depletions_2014}.}

%(!!! Add a sentence: our AGN model use a xx depletion factor, but this could change in dusty regions, like some high-z galaxies)

The O~III]/He~II ratio (Figure~\ref{fig:1}f) can also be used as metallicity diagnostics, despite being strongly affected by the ionization parameter and the $E_{\rm peak}$ (O~III]/He~II varies up to$\sim1.0\,$dex for $-3.8<\log(U)<-2.2$ and $-1.5<E_{\rm peak}<-1.0$) and depends moderately on the gas pressure (O~III]/He~II varies up to $\sim0.4$ dex for $6.2<\log(P/k)<7.8$). Additional line ratios are needed as the O~III]/He~II ratio is double-value with metallicity.

The most ideal UV metallicity diagnostics at $12+\log(\rm O/H)>8.0$ are the [N~II]$\,\lambda2143$/[O~II]$\,\lambda2470$ ratio and the N~III]$\,\lambda1747-1752$/O~III]$\,\lambda\lambda1660,6$ ratio (Figure~\ref{fig:1}g,h), as it has very little dependence on the ionization parameter, $E_{\rm peak}$, and gas pressure ([N~II]/[O~II] and N~III]/O~III] varies $\lesssim0.2\,$dex for $-3.8<\log(U)<-2.2$, $-1.5<E_{\rm peak}<-1.0$, and $6.2<\log(P/k)<7.8$). As shown in Figure~\ref{fig:B1}, the only caveat for using [N~II]/[O~II] or N~III]/O~III] to measure metallicity is their dependence on nitrogen abundance, which can vary the metallicity by up to $\sim0.4$ dex in metal-rich galaxies ($12+\log(\rm O/H)>8.5$). {\color{black}The [N~II]/[O~II] and N~III]/O~III] ratios are sensitive to metallicity at $12+\log(\rm O/H)\gtrsim8.0$ because the secondary nucleosynthetic for nitrogen dominates over the primary nucleosynthetic in this regime, which contributes to a rapidly increasing nitrogen to oxygen abundance ratio with increasing metallicity.}

%% Include any \tablenotetext{key}{text}, \tablerefs{ref list},
%% or \tablecomments{text} between the \enddata and 
%% \end{deluxetable} commands
%\tablecomments{Tricubic surface fit:$Z=a_1X+a_2Y+a_3Z+a_4X^2+a_5XY+a_6XZ+a_7Y^2+a_8YZ+a_9Z^2+a_{10}X^3+a_{11}X^2Y+a_{12}X^2Z+a_{13}XY^2+a_{14}XYZ+a_{15}XZ^2+a_{16}Y^3+a_{17}Y^2Z+a_{18}YZ^2+a_{19}Z^3+A$,\\
%where $Z=\log(R)$, $X=\log(U)$, $Y=\log(E_{peak}/keV)$}
%% No \tablerefs indicated
\subsection{Optical Metallicity Diagnostics}
\begin{figure*}[htb]
\epsscale{1.0}
\plotone{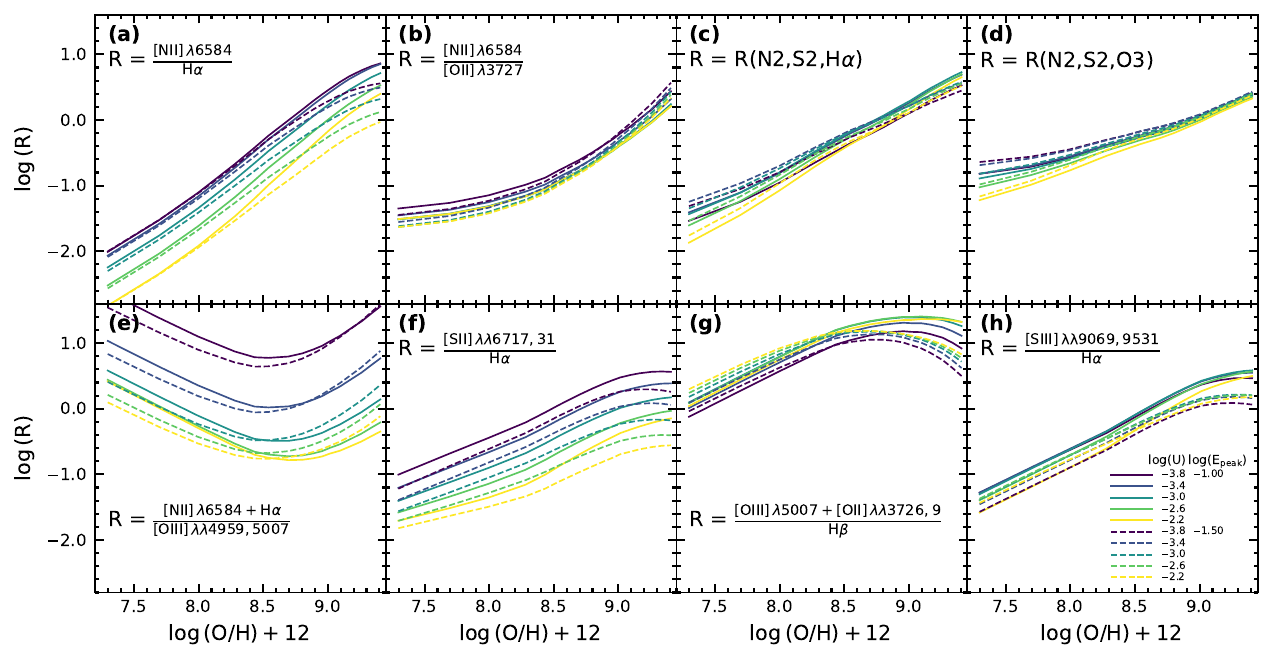}
\caption{Optical metallicity diagnostics for the AGN narrow line regions as predicted by AGN models with $\log(P/k)=7.0$, 'NHlow' nitrogen scaling relation, and varying ionization parameters ($\log(U)=-3.8,-3.4,-3.0,-2.6,-2.2$ correspond to colors from dark to bright) and varying peak energy in the AGN radiation field ($\log(E_{\rm peak}/\rm keV)=-1.0$ in solid lines and $\log(E_{\rm peak}/\rm keV)=-1.5$ in dash lines). 
\label{fig:2}}
\end{figure*}

\begin{figure*}[htb]
\epsscale{1.0}
\plotone{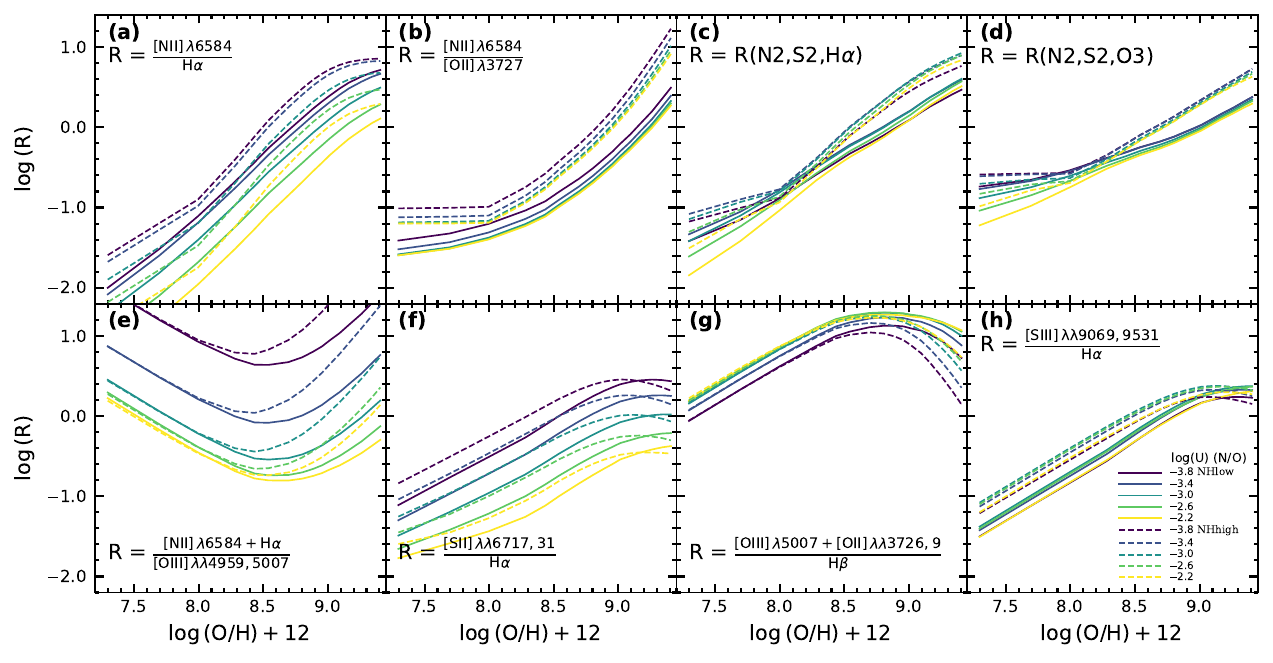}
\caption{{\color{black}The effect of nitrogen scaling relation on optical metallicity diagnostics for the AGN narrow line regions as predicted by AGN models with $\log(P/k)=7.0$ and $\log(E_{\rm peak}/\rm keV)=-1.5$.} As in Figure~\ref{fig:2}, but with line styles representing the nitrogen scaling relation used in the AGN models (`NHlow' in solid lines and `NHhigh' in dash lines). 
\label{fig:B2}}
\end{figure*}

Optical metallicity diagnostics use the brightest emission lines in optical band, including [O~II]$\,\lambda\lambda3727,9$, [O~III]$\,\lambda5007$,  [N~II]$\,\lambda6584$, [S~II]$\,\lambda\lambda6717,31$, [S~III]$\,\lambda\lambda9069,9531$, and two hydrogen lines as the baselines H$\alpha$ and H$\beta$. 

Figure~\ref{fig:2} shows the most promising optical metallicity diagnostics in NLRs derived by AGN models with $\log(P/k)=7.0$ and 'NHlow' nitrogen scaling relation, with colors and line styles corresponding to ionization parameters and $E_{\rm peak}$ of AGN radiation fields, respectively. The coefficient fits for these diagnostics are shown in Table~\ref{tab:2}. {\color{black} The coefficient fits for the same set of diagnostics but derived from AGN models with `NHhigh' nitrogen scaling relation can be found in Table~\ref{tab:C2}.}

The [N~II]$\,\lambda6584$/[O~II]$\,\lambda\lambda3727,9$ ratio (Figure~\ref{fig:2}b) is an ideal metallicity diagnostic at $12+\log(\rm O/H)\gtrsim8.0$ (where [N~II]/[O~II]$\gtrsim-1.2$) in the optical band, showing almost no dependence (metallicity varies $\lesssim0.2\,$dex at constant line ratio) on ionization parameter, $E_{\rm peak}$, and gas pressure. {\color{black}This ratio is sensitive to metallicity because of the secondary nucleosynthetic origin of nitrogen and the temperature sensitivity of the oxygen lines.}

Another nitrogen line-based metallicity diagnostic, [N~II]$\,\lambda6584$/H$\alpha$ (Figure~\ref{fig:2}a), depends strongly on ionization parameter at the full range of metallicity ([N~II]/H$\alpha$ varies up to $\sim0.6\,$dex for $-3.8<\log(U)<-2.2$) and also depends on $E_{\rm peak}$ at $12+\log(\rm O/H)>8.5$ ([N~II]/H$\alpha$ varies up to $\sim0.2\,$dex for $-1.5<E_{\rm peak}<-1.0$). Prior knowledge of the ionization parameter and $E_{\rm peak}$ is required to use [N~II]/H$\alpha$ for metallicity measurement. {\color{black}The [N~II]/H$\alpha$ ratio is sensitive to metallicity because it scales with nitrogen abundance when the nebula is hot at $12+\log(\rm O/H)\lesssim9.0$. However, at higher metallicity, nitrogen becomes a main coolant in nebulae and cools down the electronic temperature, which in turn weakens the [N~II] emission and causes the plateau of [N~II]/H$\alpha$ ratio at $12+\log(\rm O/H)\gtrsim9.1$.}

First proposed by \citet{dopita_chemical_2016} for star-forming regions, R(N2,\,S2,\,H$\alpha$) (Figure~\ref{fig:2}c) is a composite diagnostic using both [N~II]$\,\lambda6584$/H$\alpha$ and [N~II]$\,\lambda6584$/[S~II]$\,\lambda\lambda6717,31$ ratios through $\log \rm R=\log([N~II]/[S~II])+0.264\log([N~II]/H\alpha)$ to overcome the ionization parameter degeneracy. In the AGN model, this composite diagnostic also presents a dependence on the ionization parameter (R(N2,\,S2,\,H$\alpha$) varies up to $\sim0.3\,$dex for $-3.8<\log(U)<-2.2$), as well as a small dependence on the $E_{\rm peak}$. {\color{black}The [N~II]/[S~II] ratio is included in this diagnostic as it is sensitive to metallicity at $12+\log(\rm O/H)\gtrsim8.0$. The different nucleosynthetic status between nitrogen and sulfur cause the relative abundance between nitrogen and sulfur to scale with metallicity.}

If [O~III]$\,\lambda5007$/H$\beta$ is available, one can further reduce the parameter degeneracy by using R(N2, S2, O3) (Figure~\ref{fig:2}d) through $\log \rm R=0.9738\log([N~II]/[S~II])+0.047\log([N~II]/H\alpha)-0.183\log([O~III]/H\beta)$. By introducing [O~III]/H$\beta$, this new composite diagnostic has a less than $\sim0.1\,$dex scatter at the metal-rich end ($12+\log(\rm O/H)>8.5$, where R(N2, S2, O3)$\gtrsim-$0.2).

In the case when spectral resolution is not high enough to separate [N~II] and H$\alpha$, the ([N~II]+H$\alpha$)/[O~III]$\,\lambda\lambda4969,5007$ ratio can be used as metallicity diagnostic with the prior knowledge of the ionization parameter (Heretz et al. 2024, in prep). Despite the large dependence on the ionization parameter, the ([N~II]+H$\alpha$)/[O~III] ratio (Figure~\ref{fig:2}e) is almost independent of the $E_{\rm peak}$ and gas pressure (([N~II]+H$\alpha$)/[O~III] varies up to $\sim$0.1 dex for $-1.5<E_{\rm peak}<-1.0$ and $6.2<\log(P/k)<7.8$) , especially at the high metallicity regime ($12+\log(\rm O/H)>8.5$). {\color{black}Nevertheless, as a double-valued metallicity diagnostic, two metallicities will be obtained for the same ([N~II]+H$\alpha$)/[O~III] ratio. An additional metallicity diagnostic is required to determine whether the metallicity is on the upper or lower branches.} 

\begin{deluxetable*}{c|c|c|c|c|c|c|c|c}[htb]
\tabletypesize{\scriptsize}
\tablewidth{1pc}
\tablecaption{Cubic Surface Fits for Optical Metallicity Diagnostics for AGN Narrow Line Regions with $\log(P/k)=7.0$ and `NHlow' Scaling Relation\label{tab:2}}
\tablenum{3}
\tablehead{\multicolumn{9}{c}{$F=a_1X+a_2Y+a_3Z+a_4X^2+a_5XY+a_6XZ+a_7Y^2+a_8YZ+a_9Z^2+a_{10}X^3+a_{11}X^2Y$}\\
\multicolumn{9}{c}{$+a_{12}X^2Z+a_{13}XY^2+a_{14}XYZ+a_{15}XZ^2+a_{16}Y^3+a_{17}Y^2Z+a_{18}YZ^2+a_{19}Z^3+a_{20}$,}\\ 
\multicolumn{9}{c}{where $X=\log(U)$, $Y=\log(E_{\rm peak}/\rm keV)$, $Z=\log(R)$, $F=\log(\rm O/H)+12$,$-3.8\leq X\leq-1.0$, $-2.0 \leq Y\leq-1.0$}\\
\hline
\colhead{$\log$(R)} & \colhead{$\rm \frac{[NII]\,\lambda6584}{H\alpha}$} & \colhead{$\rm \frac{[NII]\,\lambda6584}{[OII]\,\lambda\lambda3727,9}$} & \colhead{\scriptsize$\rm R(N2,S2,{H\alpha})$\tablenotemark{\scriptsize a}} & \colhead{\scriptsize$\rm R(N2,S2,O3)$\tablenotemark{\scriptsize b}} & \colhead{$\rm \frac{[NII]\,\lambda6584+H\alpha}{[OIII]\,\lambda\lambda4969,5007}$} & \colhead{$\rm \frac{[SII]\,\lambda\lambda6717,31}{H\alpha}$} & \colhead{$\rm \frac{[OIII]\,\lambda5007+[OII]\,\lambda\lambda3726,9}{H\beta}$} & \colhead{$\rm \frac{[SIII]\,\lambda\lambda9069,9531}{H\alpha}$}}
\startdata
$F_{\rm min}$ & 7.30 & 7.70 & 7.30 & 7.30 & 8.54 & 7.30 & 7.30 & 7.30 \\
$F_{\rm max}$ & 9.41 & 9.41 & 9.41 & 9.41 & 9.41 & 9.09 & 8.43 & 8.96 \\
\hline
$a_1$ & -1.8782 & -0.1240 & -1.3658 & -0.0112 & -1.2871 & -1.9915 & -0.3703 & -1.2203 \\
$a_2$ & -2.0750 & -0.9102 & -4.1484 & -3.6279 & -3.2465 & 0.8524 & -0.3787 & -1.4738 \\
$a_3$ & -0.0036 & 1.1341 & 0.1233 & 1.3180 & 0.9479 & 2.3005 & 2.8720 & -0.3800 \\
$a_4$ & -0.7073 & 0.0348 & -0.4951 & 0.0139 & 0.0621 & -0.7945 & -0.3138 & -0.5075 \\
$a_5$ & -0.3310 & -0.2006 & -0.8702 & -0.4994 & -1.6210 & 0.0902 & 0.5403 & -0.5715 \\
$a_6$ & -0.2808 & 0.2457 & 0.0583 & 0.8710 & 0.1984 & 0.2329 & 0.1863 & -0.7064 \\
$a_7$ & -0.7203 & -0.3925 & -1.8512 & -1.6211 & -0.4662 & 0.8467 & -1.1201 & -0.5687 \\
$a_8$ & -0.2692 & 0.2901 & -0.8455 & -0.8937 & -0.3288 & 0.4348 & 2.4970 & 0.1039 \\
$a_9$ & 0.1196 & 0.0322 & 0.0256 & 0.2009 & -0.3891 & 1.8125 & -0.4653 & -0.4168 \\
$a_{10}$ & -0.0872 & 0.0137 & -0.0710 & -0.0054 & 0.0886 & -0.0848 & -0.0416 & -0.0626 \\
$a_{11}$ & 0.0198 & -0.0268 & -0.0615 & -0.0257 & -0.2103 & 0.0600 & -0.0099 & -0.0785 \\
$a_{12}$ & -0.0491 & 0.0427 & 0.0282 & 0.1779 & 0.0159 & -0.0252 & 0.0466 & -0.0819 \\
$a_{13}$ & -0.1185 & -0.0164 & -0.1720 & -0.0816 & -0.1972 & -0.0378 & 0.1576 & -0.1195 \\
$a_{14}$ & 0.0003 & 0.0080 & -0.0024 & 0.0864 & 0.0714 & -0.0308 & 0.0676 & -0.0989 \\
$a_{15}$ & 0.0200 & 0.0541 & 0.0393 & 0.2905 & -0.1086 & 0.4513 & 0.1494 & -0.0727 \\
$a_{16}$ & -0.0860 & -0.0447 & -0.2660 & -0.2033 & 0.1704 & 0.1411 & -0.3686 & -0.0882 \\
$a_{17}$ & 0.0398 & 0.0451 & -0.1635 & -0.2532 & -0.2006 & 0.4614 & 0.6273 & 0.3453 \\
$a_{18}$ & -0.0717 & -0.0825 & -0.1014 & -0.1145 & 0.0791 & -0.4321 & -0.7958 & -0.1910 \\
$a_{19}$ & 0.0484 & 0.2108 & 0.0208 & -0.0768 & 0.0244 & 0.6601 & 0.0533 & 0.0230 \\
$a_{20}$ & 6.5939 & 8.6815 & 6.1573 & 7.4503 & 7.0226 & 8.1778 & 6.9703 & 7.4864 \\
\hline
RMS err(\%) & 3.84\% & 3.02\% & 3.25\% & 5.65\% & 3.20\% & 6.04\% & 3.35\% & 4.03\% \\
%R2_1 & 0.9957 & 0.9961 & 0.9969 & 0.9908 & 0.9728 & 0.9798 & 0.9916 & 0.9940 \\
%R2_2 & 0.9346 & 0.9375 & 0.9447 & 0.9038 & 0.8352 & 0.8579 & 0.9081 & 0.9224 \\
%score & 0.9957 & 0.9961 & 0.9969 & 0.9908 & 0.9728 & 0.9798 & 0.9916 & 0.9940 \\
\enddata
\tablenotetext{\tiny a}{$\log \rm R(N2,S2,{H\alpha})=\log([N~II]\,\lambda6584/[S~II]\,\lambda\lambda6717,31)+0.264\log([N~II]\,\lambda6584/H\alpha)$}
\tablenotetext{\tiny b}{$\log \rm R(N2,S2,O3)=0.9738\log([N~II]\,\lambda6584/[S~II]\,\lambda\lambda6717,31)+0.047\log([N~II]\,\lambda6584/H\alpha)-0.183\log([O~III]\,\lambda5007/H\beta)$}
\end{deluxetable*}
\begin{deluxetable*}{c|c|c|c|c|c|c|c|c}[htb]
\tabletypesize{\scriptsize}
\tablewidth{1pc}
\tablecaption{Cubic Surface Fits for Optical Metallicity Diagnostics for AGN Narrow Line Regions with $\log(P/k)=7.0$ and `NHhigh' Scaling Relation\label{tab:C2}}
\tablenum{4}
\tablehead{\multicolumn{9}{c}{$F=a_1X+a_2Y+a_3Z+a_4X^2+a_5XY+a_6XZ+a_7Y^2+a_8YZ+a_9Z^2+a_{10}X^3+a_{11}X^2Y$}\\
\multicolumn{9}{c}{$+a_{12}X^2Z+a_{13}XY^2+a_{14}XYZ+a_{15}XZ^2+a_{16}Y^3+a_{17}Y^2Z+a_{18}YZ^2+a_{19}Z^3+a_{20}$,}\\ 
\multicolumn{9}{c}{where $X=\log(U)$, $Y=\log(E_{\rm peak}/\rm keV)$, $Z=\log(R)$, $F=\log(\rm O/H)+12$,$-3.8\leq X\leq-1.0$, $-2.0 \leq Y\leq-1.0$}\\
\hline
\colhead{$\log$(R)} & \colhead{$\rm \frac{[NII]\,\lambda6584}{H\alpha}$} & \colhead{$\rm \frac{[NII]\,\lambda6584}{[OII]\,\lambda\lambda3727,9}$} & \colhead{\scriptsize$\rm R(N2,S2,{H\alpha})$\tablenotemark{\scriptsize a}} & \colhead{\scriptsize$\rm R(N2,S2,O3)$\tablenotemark{\scriptsize b}} & \colhead{$\rm \frac{[NII]\,\lambda6584+H\alpha}{[OIII]\,\lambda\lambda4969,5007}$} & \colhead{$\rm \frac{[SII]\,\lambda\lambda6717,31}{H\alpha}$} & \colhead{$\rm \frac{[OIII]\,\lambda5007+[OII]\,\lambda\lambda3726,9}{H\beta}$} & \colhead{$\rm \frac{[SIII]\,\lambda\lambda9069,9531}{H\alpha}$}}
\startdata
$F_{\rm min}$ & 7.30 & 8.00 & 8.00 & 8.00 & 8.54 & 7.30 & 7.30 & 7.30 \\
$F_{\rm max}$ & 9.41 & 9.41 & 9.41 & 9.41 & 9.41 & 8.89 & 8.43 & 8.96 \\
\hline
$a_1$ & -1.9214 & -0.2611 & -0.9015 & -0.0016 & -0.4663 & -1.7941 & -0.6120 & -1.5435 \\
$a_2$ & -1.4690 & -0.2699 & -3.9591 & -3.4183 & -1.5370 & 1.6688 & -6.4836 & -2.6544 \\
$a_3$ & 0.3797 & 0.8840 & -0.4376 & 0.1725 & 0.6603 & 4.6814 & 3.7552 & 0.0963 \\
$a_4$ & -0.7238 & -0.0210 & -0.3034 & 0.0930 & 0.2449 & -0.9917 & -0.3266 & -0.6462 \\
$a_5$ & -0.4805 & -0.1973 & -0.7151 & -0.5766 & -0.8823 & 0.3830 & 0.1652 & -0.8984 \\
$a_6$ & -0.0148 & 0.0781 & 0.0906 & 0.2604 & 0.1592 & 0.9306 & 0.2108 & -0.5312 \\
$a_7$ & -0.2695 & 0.0841 & -2.0275 & -1.6740 & 0.1955 & 0.8877 & -4.2849 & -1.3093 \\
$a_8$ & -0.1845 & 0.2211 & -1.5281 & -1.5745 & -0.0304 & 1.5509 & 5.9442 & 0.5018 \\
$a_9$ & 0.3487 & -0.1150 & -0.0400 & 0.1581 & -0.3640 & 2.5431 & 1.0725 & -0.7345 \\
$a_{10}$ & -0.0881 & 0.0067 & -0.0444 & 0.0141 & 0.0989 & -0.1236 & -0.0389 & -0.0837 \\
$a_{11}$ & -0.0205 & -0.0240 & -0.0440 & -0.0391 & -0.1269 & 0.0803 & -0.0240 & -0.1015 \\
$a_{12}$ & -0.0043 & 0.0144 & 0.0485 & 0.0834 & 0.0511 & 0.0218 & 0.0824 & -0.0801 \\
$a_{13}$ & -0.0909 & -0.0197 & -0.1652 & -0.1238 & -0.0576 & -0.0114 & 0.0570 & -0.2213 \\
$a_{14}$ & -0.0308 & 0.0073 & -0.1025 & -0.1280 & 0.0210 & 0.1354 & 0.1411 & 0.0090 \\
$a_{15}$ & 0.1010 & 0.0186 & 0.0093 & 0.0474 & -0.0628 & 0.6305 & 0.2968 & -0.1133 \\
$a_{16}$ & -0.0099 & 0.0616 & -0.3307 & -0.2330 & 0.2260 & 0.1223 & -0.8766 & -0.2311 \\
$a_{17}$ & 0.0860 & 0.0306 & -0.3383 & -0.3681 & -0.0037 & 0.6003 & 1.6241 & 0.5750 \\
$a_{18}$ & -0.0719 & -0.0177 & -0.1564 & 0.0323 & 0.0430 & -0.2850 & -1.2517 & -0.5665 \\
$a_{19}$ & 0.0925 & 0.0448 & 0.0796 & -0.0620 & 0.0300 & 0.6841 & -0.6102 & 0.0957 \\
$a_{20}$ & 6.7528 & 8.5448 & 6.3420 & 7.2357 & 7.8560 & 9.4790 & 3.8901 & 6.8183 \\
\hline
RMS err(\%) & 3.59\% & 1.19\% & 3.22\% & 3.50\% & 2.66\% & 5.34\% & 3.12\% & 4.44\% \\
%R2_1 & 0.9956 & 0.9991 & 0.9935 & 0.9923 & 0.9820 & 0.9751 & 0.9830 & 0.9867 \\
%R2_2 & 0.9335 & 0.9702 & 0.9191 & 0.9122 & 0.8657 & 0.8422 & 0.8696 & 0.8845 \\
%score & 0.9956 & 0.9991 & 0.9935 & 0.9923 & 0.9820 & 0.9751 & 0.9830 & 0.9867 \\
\enddata
\tablenotetext{\tiny a}{$\log \rm R(N2,S2,{H\alpha})=\log([N~II]\,\lambda6584/[S~II]\,\lambda\lambda6717,31)+0.264\log([N~II]\,\lambda6584/H\alpha)$}
\tablenotetext{\tiny b}{$\log \rm R(N2,S2,O3)=0.9738\log([N~II]\,\lambda6584/[S~II]\,\lambda\lambda6717,31)+0.047\log([N~II]\,\lambda6584/H\alpha)-0.183\log([O~III]\,\lambda5007/H\beta)$}
\end{deluxetable*}
All optical metallicity diagnostics listed above involve the nitrogen line [N~II]$\,\lambda6584$, making them sensitive to the nitrogen abundance, which is characterized by an empirical nitrogen scaling relation in the AGN model. In reality, the nitrogen abundance in galaxies may deviate from the empirical relation due to variations in the star formation history \citep{nomoto_nucleosynthesis_2013,kobayashi_origin_2020}. In Figure~\ref{fig:B2}, we present how the change in nitrogen scaling relation affects these metallicity diagnostics. The [N~II]/H$\alpha$, [N~II]/[O~II], and R(N2, S2, H$\alpha$) ratios are affected the most, with the metallicity varying by up to $\sim0.4\,$dex at the metallicities above $12+\log(\rm O/H)>8.7$. The ([N~II]+H$\alpha$)/[O~III] and R(N2, S2, O3) ratios are less affected by the nitrogen scaling relation, with variation within $\sim0.3\,$dex at the metallicities above $12+\log(\rm O/H)>8.7$.

Using the [S~II]$\,\lambda\lambda6717,31$/H$\alpha$ ratio (Figure~\ref{fig:2}f) to estimate metallicity overcomes the nitrogen abundance dependence. With prior knowledge of the ionization parameter, the [S~II]/H$\alpha$ ratio only slightly depends on $E_{\rm peak}$ ([S~II]/H$\alpha$ varies up to $\sim0.2\,$dex for $-1.5<E_{\rm peak}<-1.0$) and is almost independent of gas pressure and nitrogen scaling relation ([S~II]/H$\alpha$ varies $\sim0.1\,$dex for $6.2<\log(P/k)<7.8$ and `NHlow'$<$N/O$<$`NHhigh').

When [S~III]$\,\lambda\lambda9069,9531$ are observed, the [S~III]/H$\alpha$ ratio (Figure~\ref{fig:2}h) is a better metallicity diagnostic than [S~II]$\,\lambda\lambda6717,31$/H$\alpha$, because the [S~III]/H$\alpha$ ratio is almost independent of the change in ionization parameter and pressure, with only small dependence on the $E_{\rm peak}$ and the nitrogen abundance scaling relation ([S~III]/H$\alpha$ varies up to $\sim0.2\,$dex for $-1.5<E_{\rm peak}<-1.0$ and `NHlow'$<$N/O$<$`NHhigh'). 

{\color{black}The [S~II]/H$\alpha$ and [S~III]/H$\alpha$ ratios are sensitive to metallicity because they scale with the sulfur abundance. Similar to [N~II]/H$\alpha$, the plateau of these line ratios at high metallicity is due to the decrease in nebular temperature, which weakens the [S~II] and [S~II] emission.}

For high-redshift galaxies where [N~II] and H$\alpha$ are beyond the observational wavelength window, an alternative metallicity diagnostic is ([O~III]$\,\lambda5007$+[O~II]$\,\lambda\lambda3727,9$)/H$\beta$ (Figure~\ref{fig:2}g). This ([O~III]+[O~II])/H$\beta$ ratio is a sensitive AGN metallicity diagnostic at $12+\log(\rm O/H)<8.5$ (where ([O~III]+[O~II])/H$\beta$$\lesssim$1.0). In this regime, the line ratio is almost independent of the ionization parameter, $E_{\rm peak}$, and gas pressure ([O~III]+[O~II])/H$\beta$ varies $\lesssim0.1\,$dex for $-3.8<\log(U)<-2.2$, $-1.5<E_{\rm peak}<-1.0$, and $6.2<\log(P/k)<7.8$). The ([O~III]+[O~II])/H$\beta$ ratio is also not affected by the nitrogen scaling relation across its valid range, making it an ideal optical metallicity diagnostic for NLRs in high-redshift metal-poor galaxies. {\color{black}However, at ([O~III]+[O~II])/H$\beta$$\gtrsim$0.4, two metallicities will be obtained for the same [O~III]+[O~II])/H$\beta$ ratio. To solve the metallicity degeneracy, an additional metallicity diagnostic is required to determine whether the metallicity is on the upper or lower branches.}
\section{ionization parameter diagnostics}
	\label{sec:ioni}

\begin{figure*}[htb]
\epsscale{1.0}
\plotone{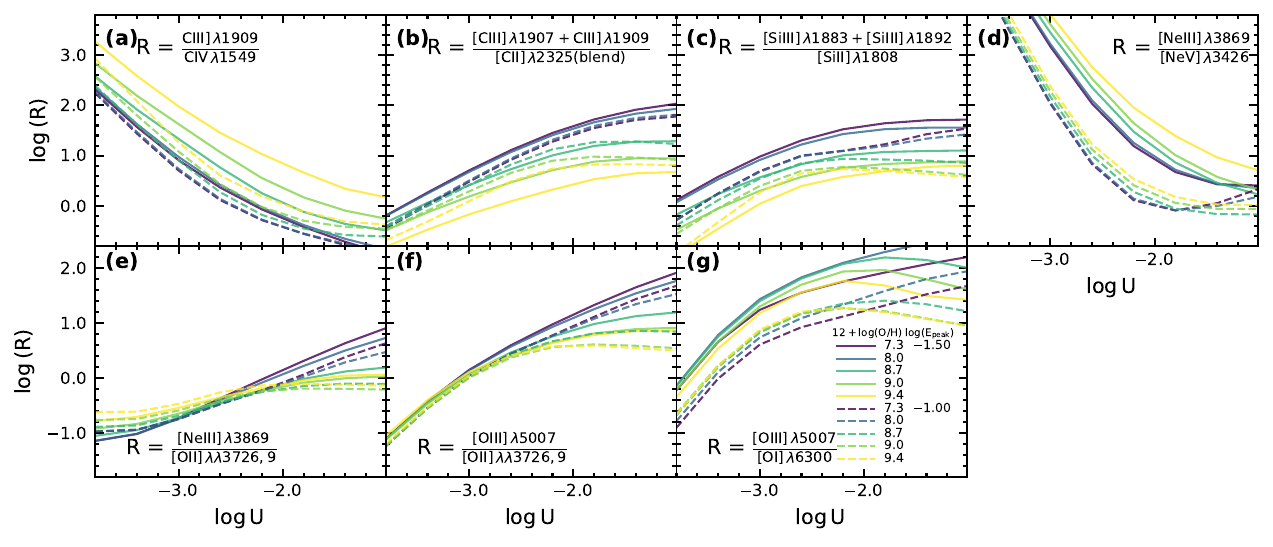}
\caption{UV and optical ionization parameter diagnostics for the AGN narrow line regions as predicted by AGN models with $\log(P/k)=7.0$, 'NHlow' nitrogen scaling relation, and varying metallicity ($12+\log(\rm O/H)=7.3,8.0,8.7,9.0,9.4$ correspond to colors from dark to bright) and varying peak energy in the AGN radiation field ($\log(E_{\rm peak}/\rm keV)=-1.0$ in solid lines and $\log(E_{\rm peak}/\rm keV)=-1.5$ in dash lines).
\label{fig:4}}
\end{figure*}

This section presents ionization parameter diagnostics for the NLRs in AGNs at UV and optical wavelengths. Ionization parameter diagnostics typically comprise two emission lines from the different ionization states of the same atomic species (i.e., argon, carbon, sulfur, silicon, neon, nitrogen, and oxygen). The ionization-sensitive line ratios from different species can also be used as ionization parameter diagnostics. 

As shown in Appendix B Figure~\ref{fig:B4}, most ionization parameter diagnostics are insensitive to the change in nitrogen scaling relations in AGN models. Therefore, the reliability of ionization parameters diagnostics is mainly affected by metallicity, $E_{peak}$, and gas pressure. 

Figure~\ref{fig:4} shows the most promising ionization parameter diagnostics in NLRS derived from AGN models with $\log(P/k)=7.0$ and 'NHlow' nitrogen scaling relation, covering the UV and optical wavelengths. The colors and line styles correspond to metallicity and $E_{peak}$, respectively. The coefficient fits for these diagnostics are shown in Table~\ref{tab:4}.

As stated in Section 2.2, the ionization parameter $U$ in our models represents the dimensionless ionization parameter at the inner edge of the cloud. This has an important effect that at $\log(U)\gtrsim-2.0$, when radiation pressure dominates over the gas pressure, the local ionization parameter in the cloud becomes insensitive to the initial ionization parameter at the inner edge of the cloud  \citep{dopita_are_2002,groves_dusty_2004-1}. In this regime, ionization parameter diagnostics become insensitive to the initial ionization parameter due to the decoupling of the local ionization parameter and initial ionization parameter. We discuss the valid range of each ionization parameter diagnostic separately below. {\color{black} All emission line ratios discussed in this section are in log scale.}

\begin{deluxetable*}{c|c|c|c|c|c|c|c}
\tabletypesize{\scriptsize}
\tablewidth{1pc}
\tablecaption{Cubic Surface Fits for Ionization Parameter Diagnostics for AGN Narrow Line Regions with $\log(P/k)=7.0$ and `NHlow' Scaling Relation\label{tab:4}}
\tablenum{5}
\tablehead{\multicolumn{8}{c}{$F=a_1X+a_2Y+a_3Z+a_4X^2+a_5XY+a_6XZ+a_7Y^2+a_8YZ+a_9Z^2+a_{10}X^3+a_{11}X^2Y$}\\
\multicolumn{8}{c}{$+a_{12}X^2Z+a_{13}XY^2+a_{14}XYZ+a_{15}XZ^2+a_{16}Y^3+a_{17}Y^2Z+a_{18}YZ^2+a_{19}Z^3+a_{20}$,}\\ 
\multicolumn{8}{c}{where $X=\log(\rm O/H)+12$, $Y=\log(E_{\rm peak}/\rm keV)$, $Z=\log(R)$, $F=\log(U)$, $7.29\leq X\leq9.41$, $-2.0 \leq Y\leq-1.0$}\\ 
\hline
\colhead{$\log$(R)} & \colhead{$\rm \frac{CIII]\,\lambda1909}{CIV\,\lambda1549}$} & \colhead{$\rm\frac{[CIII]\,\lambda1907+CIII]\,\lambda1909}{[CII]\,\lambda2325(blend)}$} & \colhead{$\rm \frac{[SiIII]\,\lambda\lambda1883,92}{[SiII]\,\lambda1808}$} & \colhead{$\rm \frac{[NeIII]\,\lambda3869}{[NeV]\,\lambda3426}$} & \colhead{$\rm \frac{[NeIII]\,\lambda3869}{[OII]\,\lambda\lambda3727,9}$} & \colhead{$\rm \frac{[OIII]\,\lambda5007}{[OII]\,\lambda\lambda3727,9}$} & \colhead{$\rm \frac{[OIII]\,\lambda5007}{[OI]\,\lambda6300}$}}
\startdata
$F_{\rm min}$ & -3.8 & -3.8 & -3.8 & -3.8 & -3.4 & -3.8 & -3.8 \\
$F_{\rm max}$ & -1.8 & -2.4 & -2.4 & -2.2 & -2.2 & -2.4 & -2.4 \\
\hline
$a_1$ & 17.5531 & 36.3805 & -13.4037 & 2.4478 & -16.8483 & -12.7880 & -5.4719 \\
$a_2$ & -19.6169 & -39.7556 & -7.0918 & 4.9591 & -25.7719 & 4.2263 & 9.4866  \\
$a_3$ & -5.8276 & 9.2755 & 1.2417 & -1.4096 & 16.1666 & 2.2411 & 9.2675 \\
$a_4$ & -2.0524 & -3.9871 & 1.7048 & -0.3011 & 2.4469 & 1.4680 & 0.6292 \\
$a_5$ & 4.6750 & 8.7311 & 4.9286 & 1.1142 & 4.6659 & -0.7580 & 0.5409 \\
$a_6$ & 1.1560 & -1.6840 & 0.3070 & 0.1783 & -3.9705 & -0.3199 & -1.6266 \\
$a_7$ & -2.5523 & -4.9786 & 7.8291 & 5.7040 & -6.5029 & -0.0403 & 7.1228 \\
$a_8$ & -1.0325 & 3.6635 & 3.5109 & -0.7430 & 1.4741 & 0.7852 & 3.0119 \\
$a_9$ & 0.0569 & 0.0360 & -0.2185 & 0.0147 & -1.8897 & -0.2470 & 0.4681 \\
$a_{10}$ & 0.0783 & 0.1454 & -0.0742 & 0.0118 & -0.1136 & -0.0569 & -0.0271 \\
$a_{11}$ & -0.3058 & -0.4890 & -0.3449 & -0.0762 & -0.2636 & 0.0259 & -0.0834 \\
$a_{12}$ & -0.0688 & 0.0851 & -0.0378 & -0.0104 & 0.2745 & 0.0204 & 0.0858 \\
$a_{13}$ & 0.0696 & 0.3908 & -0.1424 & 0.0646 & 0.2184 & -0.0967 & -0.2624 \\
$a_{14}$ & 0.1002 & -0.3738 & -0.3474 & 0.0499 & -0.0714 & -0.0615 & -0.1462 \\
$a_{15}$ & 0.0245 & -0.0094 & 0.0433 & 0.0058 & 0.1829 & 0.0658 & -0.0157 \\
$a_{16}$ & -0.7610 & -0.6297 & 1.4752 & 1.0748 & -1.3018 & -0.3489 & 0.9876 \\
$a_{17}$ & -0.1550 & 0.2921 & 0.2883 & -0.2757 & 0.3335 & 0.0969 & 0.6019 \\
$a_{18}$ & 0.0250 & -0.0884 & -0.0387 & -0.0310 & -0.5180 & 0.0171 & 0.1975 \\
$a_{19}$ & -0.0367 & -0.0224 & 0.0230 & -0.0084 & 0.2898 & 0.1110 & 0.0675 \\
$a_{20}$ & -52.1421 & -113.6945 & 37.5155 & -4.6356 & 32.6309 & 34.7505 & 18.3828 \\
\hline
RMS err(\%) & 4.17\% & 2.91\% & 4.15\% & 1.27\% & 3.73\% & 1.64\% & 3.03\% \\
%R2_1 & 0.9961 & 0.9955 & 0.9908 & 0.9995 & 0.9930 & 0.9986 & 0.9951 \\
%R2_2 & 0.9372 & 0.9328 & 0.9040 & 0.9769 & 0.9164 & 0.9621 & 0.9299 \\
%score & 0.9961 & 0.9955 & 0.9908 & 0.9995 & 0.9930 & 0.9986 & 0.9951 \\
\enddata
\end{deluxetable*}
\subsection{UV Ionization Parameter Diagnostics}

Carbon lines in the UV wavelength provide two well-performing ionization parameter diagnostics, C~III]$\,\lambda1909$/C~IV$\,\lambda1549$ and ([C~III]$\,\lambda1907$+C~III]$\,\lambda1909$)/[C~II]$\,\lambda2325$(blend) (Figure~\ref{fig:4}a,b). {\color{black}The C~III]/C~IV ratio is sensitive to the ionization parameter at C~III]/C~IV$\gtrsim-$0.2. C~III]/[C~II] ratio is sensitive to the ionization parameter at C~III]/[C~II]$\lesssim$1.0. They are only moderately sensitive to the metallicity at $12+\log(\rm O/H)<9.0$ (C~III]/C~IV and C~III]/[C~II] varies up to $\sim0.6$\,dex for $7.3<12+\log(\rm O/H)<9.0$).} In addition, at the same metallicity range, these carbon lines ratios slightly depend on $E_{\rm peak}$ (C~III]/C~IV and C~III]/[C~II] varies up to $\sim0.4$\,dex for $-1.5<E_{\rm peak}<-1.0$) and are independent of gas pressure, making them ideal ionization parameter diagnostics over a large range of metallicity ($12+\log(\rm O/H)\lesssim9.0$).

UV emission lines from silicon can also be used to interpret ionization parameters. The [Si~III]$\,\lambda\lambda1883,92$/[Si~II]$\,\lambda1808$ ratio (Figure~\ref{fig:4}c) is sensitive to the ionization parameter {\color{black}at [Si~III]/[Si~II]$\lesssim$1.0}, while presenting limited variation to metallicity at $12+\log(\rm O/H)\lesssim$9.0. This [Si~III]/[Si~II] ratio slightly depends on the $E_{peak}$ ([Si~III]/[Si~II] varies $\lesssim0.4$\,dex for $-1.5<E_{\rm peak}<-1.0$), and is almost independent of the gas pressure ([Si~III]/[Si~II] varies $\lesssim0.1$\,dex for $6.2<\log(P/k)<7.8$). With a rough estimation of metallicity and $E_{peak}$, the ionization parameter can be estimated through the [Si~III]/[Si~II] ratio.

\subsection{Optical Ionization Parameter Diagnostics}

The most reliable optical ionization parameter diagnostic is the [O~III]$\,\lambda5007$/[O~II]$\,\lambda\lambda3727,9$ ratio (Figure~\ref{fig:4}f), which presents no dependence on metallicity, $E_{peak}$, nitrogen abundance, and only small variation with gas pressure at [O~III]/[O~II]$\lesssim$0.8 ([O~III]/[O~II] varies $\lesssim0.2\,$dex for $6.2<\log(P/k)<7.8$). At higher ionization parameters, the [O~III]/[O~II] ratio starts to vary with metallicity, $E_{peak}$, and gas pressure, becoming less sensitive to the ionization parameter, especially at metal-rich regimes ($12+\log(\rm O/H)>8.7$).

When [O~II] is beyond the observed wavelength, the [O~III]$\,\lambda5007$/[O~I]$\,\lambda6300$ ratio (Figure~\ref{fig:4}g) can be used to estimate the ionization parameter instead. Despite showing a stronger dependence on $E_{peak}$ ([O~III]/[O~I] varies up to $\sim0.8\,$dex for $-1.5<E_{\rm peak}<-1.0$), the variation of the [O~III]/[O~I] ratio on metallicity and gas pressure is similar to the [O~III]/[O~II] ratio. {\color{black}One caveat for this diagnostic is that [O~I] is produced in a partially ionized zone that lies at the edge of the nebula \citep{dopita_astrophysics_2003,groves_dusty_2004-2}. The completeness of this partially ionized zone can be affected by pressure-bounded or density-bounded nebula assumptions, potentially resulting in a weaker observed [O~I] emission line than the model prediction and an overestimation of the ionization parameter.}

On the other hand, when [O~III] is beyond the observed wavelength, neon lines can be used to estimate the ionization parameter instead. The [Ne~III]$\,\lambda3869$/[O~II]$\,\lambda\lambda3727,9$ ratio (Figure~\ref{fig:4}e) is a sensitive ionization parameter diagnostic between $-0.8\lesssim$[Ne~III]/[O~II]$\lesssim-0.2$, with small variation with metallicity ([Ne~III]/[O~II] varies up to $\sim0.4\,$dex for $7.3<12+\log(\rm O/H)<9.4$) and limited dependence on gas pressure and $E_{peak}$ ([Ne~III]/[O~II] varies $\sim0.2\,$dex for $6.2<\log(P/k)<7.8$).

The [Ne~III]$\,\lambda3869$/[Ne~V]$\,\lambda3426$ ratio (Figure~\ref{fig:4}d) is sensitive to ionization parameter at [Ne~III]/[Ne~V]$\gtrsim$0.4, with a moderate dependence on metallicity ([Ne~III]/[Ne~V] varies up to $\sim0.4$\,dex for $7.3<12+\log(\rm O/H)<9.4$) and no dependence on gas pressure at $12+\log(\rm O/H)\lesssim9.0$. Nevertheless, this ionization parameter diagnostic is strongly affected by $E_{peak}$ ([Ne~III]/[Ne~V] varies up to $\sim0.8\,$dex for $-1.5<E_{\rm peak}<-1.0$), requiring prior knowledge of $E_{peak}$ before its application.

\section{Pressure diagnostics}

\begin{figure*}[htb]
\epsscale{1.0}
\plotone{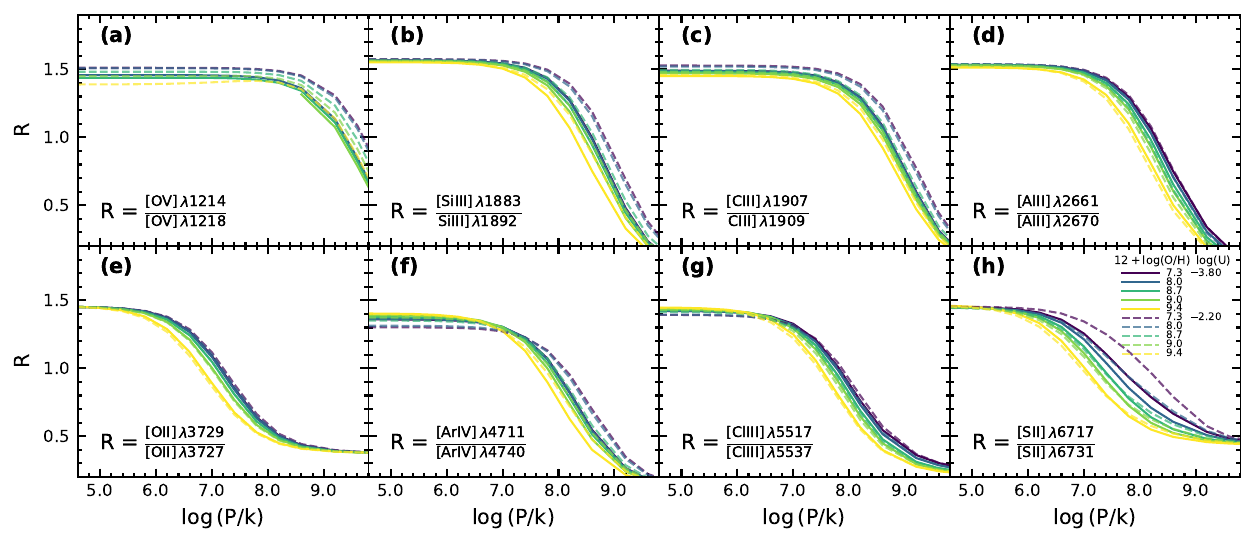}
\caption{UV and optical gas pressure diagnostics for the AGN narrow line regions as predicted by AGN models with $\log(E_{\rm peak}/\rm keV)=-1.25$, 'NHlow' nitrogen scaling relation, and varying metallicity ($12+\log(\rm O/H)=7.3,8.0,8.7,9.0,9.4$ correspond to colors from dark to bright) and varying ionization parameter ($\log(U)=-3.8$ in solid lines and $\log(U)=-2.2$ in dash lines).
\label{fig:5}}
\end{figure*}

Our AGN model assumes pressure equilibrium in the nebula, that is, the sum of radiation pressure and gas pressure remains the same within the photoionized region while the temperature and electron density vary accordingly. In this section, we present gas pressure diagnostics for AGN NLRs at UV and optical wavelengths based on our AGN model. Similar to HII regions \citep{kewley_theoretical_2019}, we find that gas pressure diagnostics have a large dependence on gas metallicity (up to an order of magnitude in $\log(P/k)$). At the same time, the effect of the ionization parameter, nitrogen abundance, and $E_{peak}$ are relatively small. Figure~\ref{fig:B5} in Appendix B shows that the change in nitrogen scaling relations barely affects these pressure diagnostics.

Figure~\ref{fig:5} presents the most promising gas pressure diagnostics in NLRs derived from AGN models with $\log(E_{\rm peak}/\rm keV)=-1.25$ and 'NHlow' nitrogen scaling relation, at UV and optical wavelengths. The colors and line styles correspond to metallicity and ionization parameters. The coefficient fits for these diagnostics are shown in Table~\ref{tab:5}. {\color{black} All emission line ratios discussed in this section are in linear scale.}

\begin{deluxetable*}{c|c|c|c|c|c|c|c|c}[htb]
\tabletypesize{\scriptsize}
\tablewidth{1pc}
\tablecaption{Cubic Surface Fits for Pressure Diagnostics for AGN Narrow Line Regions with $\log(E_{\rm peak}/\rm keV)=-1.25$ and `NHlow' Scaling Relation\label{tab:5}}
\tablenum{6}
\tablehead{\multicolumn{9}{c}{$F=a_1X+a_2Y+a_3Z+a_4X^2+a_5XY+a_6XZ+a_7Y^2+a_8YZ+a_9Z^2+a_{10}X^3+a_{11}X^2Y+a_{12}X^2Z$}\\
\multicolumn{9}{c}{$+a_{13}XY^2+a_{14}XYZ+a_{15}XZ^2+a_{16}Y^3+a_{17}Y^2Z+a_{18}YZ^2+a_{19}Z^3+a_{20}$,}\\ 
\multicolumn{9}{c}{where $X=\log(\rm O/H)+12$, $Y=\log(U)$, $Z=R$, $F=\log(P/k)$, $7.29\leq X\leq9.41$, $-3.8\leq Y\leq-1.0$}\\ 
\hline
\colhead{R} & \colhead{$\rm \frac{[OV]\,\lambda1214}{[OV]\,\lambda1218}$} & \colhead{$\rm\frac{[SiIII]\,\lambda1883}{SiIII]\,\lambda1892}$} & \colhead{$\rm \frac{[CIII]\,\lambda1907}{CIII]\,\lambda1909}$} & \colhead{$\rm \frac{[AlII]\,\lambda2661}{[AlII]\,\lambda2670}$} & \colhead{$\rm \frac{[OII]\,\lambda3729}{[OII]\,\lambda3727}$} & \colhead{$\rm \frac{[ArIV]\,\lambda4711}{[ArIV]\,\lambda4740}$} & \colhead{$\rm \frac{[ClIII]\,\lambda5517}{[ClIII]\,\lambda5537}$} & \colhead{$\rm \frac{[SII]\,\lambda6717}{[SII]\,\lambda6731}$}}
\startdata
$F_{\rm min}$ & 8.6 & 7.5 & 7.5 & 7.2 & 6.0 & 7.2 & 7.0 & 6.2 \\
$F_{\rm max}$ & 10.0 & 9.8 & 9.8 & 9.8 & 8.0 & 9.0 & 9.0 & 8.4 \\
\hline
$a_1$ & -2.8675 & -1.0783 & 1.3792 & -11.3386 & -13.7415 & -1.1142 & -9.2205 & -17.5902 \\
$a_2$ & 1.7614 & 1.4292 & 0.5959 & 1.0721 & 0.2375 & 2.3356 & 2.4467 & 2.7244 \\
$a_3$ & -17.5947 & -4.1216 & -3.6576 & -8.1055 & -11.555 & 1.7579 & -8.864 & -5.8746 \\
$a_4$ & 0.1846 & 0.0387 & -0.2292 & 1.3015 & 1.679 & 0.0967 & 1.0038 & 1.9685 \\
$a_5$ & -0.5039 & -0.7107 & -0.5491 & -0.4189 & -0.2449 & -0.7656 & -0.6898 & -0.9759 \\
$a_6$ & 2.1641 & 0.4273 & 0.3023 & 1.0232 & 0.4034 & -0.4985 & 0.4484 & -1.798 \\
$a_7$ & -0.3935 & -0.5733 & -0.6126 & -0.2599 & -0.2813 & -0.2979 & -0.3054 & -0.7498 \\
$a_8$ & -0.711 & 0.2581 & 0.5993 & 0.079 & -0.0475 & 0.4442 & -0.4881 & 0.3808 \\
$a_9$ & 8.1049 & 2.4033 & 3.5828 & 3.8642 & 9.1272 & 0.1593 & 5.5092 & 11.2749 \\
$a_{10}$ & -0.0033 & 0.0018 & 0.0116 & -0.0502 & -0.0682 & -0.0023 & -0.0369 & -0.0735 \\
$a_{11}$ & 0.0265 & 0.0467 & 0.0372 & 0.0259 & 0.0171 & 0.0427 & 0.0414 & 0.0639 \\
$a_{12}$ & -0.1037 & -0.0389 & -0.0365 & -0.0674 & -0.0439 & -0.016 & -0.0384 & 0.0481 \\
$a_{13}$ & 0.0075 & 0.0194 & 0.0185 & 0.0048 & 0.0111 & -0.0066 & 0.0097 & 0.0399 \\
$a_{14}$ & 0.0588 & -0.0187 & -0.0439 & -0.0324 & -0.0119 & -0.0274 & 0.0078 & -0.0178 \\
$a_{15}$ & -0.168 & 0.0798 & 0.0657 & -0.0098 & 0.1594 & 0.4161 & 0.1723 & 0.6055 \\
$a_{16}$ & -0.0429 & -0.0408 & -0.0468 & -0.0263 & -0.0164 & -0.0397 & -0.031 & -0.0506 \\
$a_{17}$ & -0.0129 & 0.0177 & 0.0383 & -0.0329 & -0.0027 & -0.0246 & -0.0415 & 0.0365 \\
$a_{18}$ & 0.0895 & -0.0061 & 0.0118 & 0.006 & 0.0694 & -0.2443 & 0.0818 & -0.0991 \\
$a_{19}$ & -2.662 & -1.3324 & -1.8564 & -1.6016 & -3.7644 & -2.078 & -2.729 & -5.7952 \\
$a_{20}$ & 24.9671 & 14.447 & 6.9311 & 42.6513 & 48.2778 & 12.8428 & 39.0378 & 66.4046 \\
\hline
RMS err(\%) & 1.52\% & 1.76\% & 2.08\% & 1.74\% & 1.59\% & 1.53\% & 1.64\% & 4.34\% \\
%0.999 & 0.9989 & 0.9985 & 0.999 & 0.9994 & 0.9987 & 0.9993 & 0.9958 \\
%0.968 & 0.967 & 0.961 & 0.969 & 0.975 & 0.9638 & 0.9742 & 0.9348 \\
%0.999 & 0.9989 & 0.9985 & 0.999 & 0.9994 & 0.9987 & 0.9993 & 0.9958 \\
\enddata
\end{deluxetable*}
\subsection{UV Gas Pressure Diagnostics}

At UV wavelengths, gas pressure diagnostics can be obtained by adopting the flux ratio of two emission lines from the same atom with different radiative transition probabilities. {\color{black} Emission lines with a small radiative transition probability have a low critical density and are predominantly generated by collisional de-excited, resulting in line fluxes proportional to the electron density $n_e$. On the other hand, emission-lines with large radiative transition probabilities have a high critical density and are mainly produced by radiative de-excitation when the gas density is less than the critical density, which leads to line fluxes proportional to $n_e^2$ \citep{kewley_theoretical_2019}. Therefore, the ratio of these two lines is sensitive to the electron density $n_e$ and can indicate the gas pressure.} This section discusses UV gas pressure diagnostics involving emission lines that are currently detected.

The [O~V]$\,\lambda1214$/[O~V]$\,\lambda1218$ ratio (Figure~\ref{fig:5}a) is sensitive to gas pressure in the high-pressure region {\color{black}($\log(P/k)>8.0$, where [O~V]/[O~V]$\lesssim$1.3).} The presence of these lines requires high-energy ionizing photons and is mostly detected in AGN or shock regions. This gas pressure diagnostic moderately depends on gas metallicity and the ionization parameter ([O~V]/[O~V] varies up to $\sim0.3$ for $7.3<12+\log(\rm O/H)<9.4$ and $-3.8<\log(U)<-2.2$). At this short wavelength, a broad emission line at [N~V]$\,\lambda1240$ is frequently present, making it a challenge to accurately measure the fluxes of nearby narrow lines.

The [Si~III]$\,\lambda1883$/Si~III]$\,\lambda1892$ ratio (Figure~\ref{fig:5}b) is an ideal UV pressure diagnostic first proposed by \citet{dufton_si_1984}. Compared with the [O~V]/[O~V] emission line ratio mentioned above, the [Si~III]/Si~III] ratio probes regions with medium to high pressure ($\log(P/k)>7.5$, where [Si~III]/Si~III]$\lesssim$1.5) in the nebula but is less affected by the broad emission lines and has a smaller dependence on gas metallicity and ionization parameter ([Si~III]/Si~III] varies up to $\sim0.2\,$dex for $7.3<12+\log(\rm O/H)<9.4$ and $-3.8<\log(U)<-2.2$). $E_{peak}$ only has a small impact on this gas pressure diagnostic ([Si~III]/Si~III] varies $\sim0.1$ for $-1.5<E_{\rm peak}<-1.0$) at high metallicity ($12+\log(\rm O/H)\gtrsim9.0$). 

Using the most frequently detected UV narrow emission lines, the [C~III]$\,\lambda1907$/C~III]$\,\lambda1909$ ratio (Figure~\ref{fig:5}c) provides a good measurement of gas pressure for high-pressure regions in the nebula ($\log(P/k)\gtrsim8.0$, where [C~III]/C~III]$\lesssim$1.4). Similar to the [Si~III] ratio, this [C~III] ratio has a small dependence on gas metallicity ([C~III]/C~III] varies up to $\sim0.1$ for $7.3<12+\log(\rm O/H)<9.4$) and is only slightly affected by the varied ionization parameter and the $E_{peak}$ ([C~III]/C~III] varies $\lesssim0.1$ for $-3.8<\log(U)<-2.2$ and $-1.5<E_{\rm peak}<-1.0$). The challenge of applying this pressure diagnostics lies in obtaining a high spectral resolution spectrum that can deblend these two emission lines since they are only separated in wavelength by 3$\AA$.

At longer wavelength, the [Al~II]$\,\lambda2661$/[Al~II]$\,\lambda2670$ ratio (Figure~\ref{fig:5}d) can probe gas pressure in regions with medium to high gas pressure ($\log(P/k)\gtrsim7.2$, where [Al~II]/[Al~II] $\lesssim$1.4). This ratio has a moderate dependence on gas metallicity ([Al~II]/[Al~II] varies up to $\sim0.4$ for $7.3<12+\log(\rm O/H)<9.4$) and is almost independent of the ionization parameter. Compared to other UV pressure diagnostics, this [Al~II] ratio has the strongest impact from $E_{peak}$, which, with the same line ratio, can change the gas pressure prediction by up to $\sim0.4\,$dex. 

\subsection{Optical Gas Pressure Diagnostics}

In optical wavelengths, gas pressure diagnostics are usually the ratio of two forbidden lines from the same atom species. {\color{black} These forbidden lines have small radiative transition probabilities and low critical density. At low density, the collisional de-excitation rate is comparable with the radiative de-excitation rate, resulting in the line fluxes scale with $n_e^2$ at low density and with $n_e$ at high density \citep{kewley_theoretical_2019}. Therefore, the ratio of these two forbidden lines is sensitive to the $n_e$ between the critical density of these two lines, making them pressure diagnostics in an ideal gas where pressure is directly proportional to density according to the ideal gas law.} 

A well-known example is the [O~II]$\,\lambda3729$/[O~II]$\,\lambda3727$ ratio (Figure~\ref{fig:5}e), which has been used to determine electron density in nebulae for decades \citep{osterbrock_astrophysics_1989}. In the medium-pressure ($6.2\lesssim\log(P/k)\lesssim8.0$, where 1.4$\gtrsim$[O~II]/[O~II]$\gtrsim$0.5) regions, this [O~II] line ratio is an ideal gas pressure diagnostic as it only has a small dependence on gas metallicity ([O~II]/[O~II] varies up to $\sim0.4$ for $7.3<12+\log(\rm O/H)<9.4$) and is not affected by the change in the ionization parameter and the $E_{peak}$ at $12+\log(\rm O/H)\lesssim9.0$. These two [O~II] lines have similar ionization potentials to hydrogen and are frequently observed in the optical spectra of Seyfert galaxies. However, as the two emission lines are only separated in wavelength by 2$\AA$, high-resolution spectroscopy is required to resolve the doublet.

Another pair of optical emission lines that are frequently used to measure gas pressure in the nebula is [S~II]$\,\lambda6717$ and [S~II]$\,\lambda6731$ (Figure~\ref{fig:5}h), mainly due to their wider separation in wavelength and is available in many optical observations. The [S~II]$\,\lambda6717$/[S~II]$\,\lambda6731$ ratio is sensitive to gas pressure in regions with $6.5\lesssim\log(P/k)\lesssim8.0$ (1.3$\gtrsim$[S~II]/[S~II]$\gtrsim$0.6) and has a small dependence on the ionization parameter and $E_{peak}$ ([S~II]/[S~II] varies $\lesssim0.2$ for $-3.8<\log(U)<-2.2$ and $-1.5<E_{\rm peak}<-1.0$). However, this [S~II] ratio is significantly sensitive to metallicity in the measurement of gas pressure ([S~II]/[S~II] varies $\sim0.4$ for $7.3<12+\log(\rm O/H)<9.4$). Prior knowledge of the metallicity is required to obtain reliable pressure measurements from this optical [S~II] ratio.

For nebula with high-pressure regions ($\log(P/k)\gtrsim7.5$, where [Ar~IV]/[Ar~IV]$\lesssim$1.2), the [Ar~IV]$\,\lambda4711$/[Ar~IV]$\,\lambda4740$ ratio (Figure~\ref{fig:5}f) can be used to measure the gas pressure instead. This [Ar~IV] line ratio has a small dependence on gas metallicity ([Ar~IV]/[Ar~IV] varies up to $\sim0.2$ for $7.3<12+\log(\rm O/H)<9.4$), the ionization parameter, and the $E_{peak}$ ([Ar~IV]/[Ar~IV] varies $\lesssim0.1$ for $-3.8<\log(U)<-2.2$ and $-1.5<E_{\rm peak}<-1.0$). With a large ionization potential (40.73 eV), these [Ar~IV] lines are produced in the hot regions of the nebula where the auroral [O~III]$\,\lambda4363$ line is also emitted. Therefore, this [Ar~IV] line ratio can be used to probe the gas pressure in the high-temperature, high-ionization zone of the nebula.

 Another optical gas pressure diagnostic for high-pressure regions ($\log(P/k)\gtrsim7.5$, where [Cl~III]/[Cl~III]$\lesssim$1.2) is the [Cl~III]$\,\lambda5517$/[Cl~III]$\,\lambda5537$ ratio (Figure~\ref{fig:5}g), which also has a small dependence on gas metallicity([Cl~III]/[Cl~III] varies up to $\sim0.2$ for $7.3<12+\log(\rm O/H)<9.4$). Compared to the [Ar~IV] line ratio, the pressure measured with [Cl~III] line ratio is not affected by the ionization parameter but varies $\sim0.1$ for $-1.5<E_{\rm peak}<-1.0$.

\section{$E_{\rm peak}$ diagnostics}

\begin{figure*}[bt]
\epsscale{1.0}
\plotone{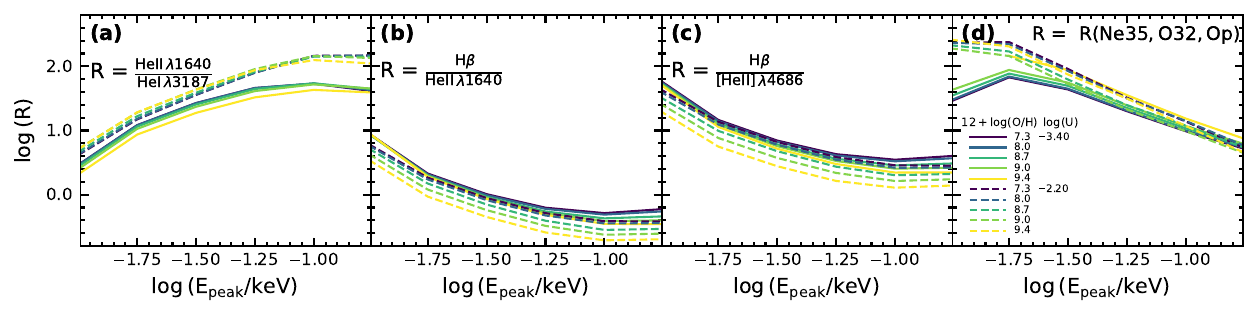}
\caption{UV and optical $E_{\rm peak}$ diagnostics for the AGN narrow line regions as predicted by AGN models with $\log(P/k)=7.0$, 'NHlow' nitrogen scaling relation, and varying metallicity ($12+\log(\rm O/H)=7.3,8.0,8.7,9.0,9.4$ correspond to colors from dark to bright) and varying ionization parameter ($\log(U)=-3.8$ in solid lines and $\log(U)=-2.2$ in dash lines).
\label{fig:6}}
\end{figure*}

$E_{\rm peak}$, the peak energy of the accretion disk emission that contributes to the big blue bump component of the radiation field of AGNs, is determined by the central black hole mass, black hole accretion rate, and the coronal radius according to the thin-disk black hole accretion model \citep{done_intrinsic_2012,thomas_physically_2016}. Therefore, the value of $E_{\rm peak}$ could vary from several eV to $\sim200\,$eV in galaxies with different black hole masses and Eddington ratios.

As discussed in the previous section, some metallicity and ionization parameter diagnostics have moderate to strong dependence on $E_{peak}$, making it necessary to estimate $E_{peak}$ before measuring metallicity and ionization parameters. In this section, we explore diagnostics for $E_{\rm peak}$ using narrow emission lines in the AGN regions, which, in the end, can also provide a unique approach to estimate the Eddington ratio or the coronal radius for actively accreting black holes. 

In Figure~\ref{fig:6}, we present four UV and optical $E_{\rm peak}$ diagnostics for NLRs derived from AGN models with $\log(P/k)=7.0$ and 'NHlow' nitrogen scaling relation. The colors and styles of the lines correspond to metallicity and the ionization parameter, respectively. We also present coefficient fits for these diagnostics in Table~\ref{tab:6}. {\color{black} All emission line ratios discussed in this section are in log scale.}
 
\subsection{$E_{\rm peak}$ Diagnostics with He~II Emission Lines}

{\color{black}With an ionization energy of 54.4 eV, the detection of He$^{+}$ via He~II emission lines is associated with the presence of a hard ionizing source \citep[e.g.][]{shirazi_strongly_2012,berg_intense_2019,tozzi_unveiling_2023}. Therefore, the line ratio between He~II emission line and a low-ionization emission line can indicate the hardness of the ionizing radiation field.} Two He~II emission lines are frequently used in previous studies, the He~II$\,\lambda1640$ and [He~II]$\,\lambda4686$ lines. 

In the UV, the He~II$\,\lambda1640$/He~I$\,\lambda3187$ ratio (Figure~\ref{fig:6}a) can serve as an $E_{\rm peak}$ diagnostic in the range of $\log(E_{\rm peak}/\rm keV)<-1.25$ {\color{black}(where He~II/He~I$\lesssim$1.6-2.0)}, with small dependence on metallicity and moderate dependence on the ionization parameter. As shown in Figure~\ref{fig:6}, at the same He~II$\,\lambda1640$/He~I$\,\lambda3187$ ratio, different metallicities can change the predicted value for $\log(E_{\rm peak})$ up to $\sim0.1\,$dex and different ionization parameters can change the prediction up to $\sim0.25\,$dex. This He~II/He~I $E_{\rm peak}$ diagnostic is unaffected by gas pressure and nitrogen abundance. The large wavelength difference between these two lines makes dust extinction correction important.

In the optical, the H$\beta$/[He~II]$\,\lambda4686$ ratio (Figure~\ref{fig:6}c) can be used to measure $E_{\rm peak}$ for soft to moderate AGN radiation fields ($\log(E_{\rm peak}/\rm keV)<-1.25$, where H$\beta$/[He~II]$\gtrsim$0.6), with no dependence on gas pressure and nitrogen abundance. At the low ionization parameters ($\log(U)\lesssim-3.0$), the H$\beta$/He~II diagnostic depends only slightly on metallicity (vary the $E_{\rm peak}$ measurement by up to $\sim0.05\,$dex). But at high ionization parameters  ($\log(U)\gtrsim-2.4$), the $E_{\rm peak}$ derived can vary with metallicity by up to $\sim0.25\,$dex. 

Across the UV and optical wavelengths, the H$\beta$/He~II$\,\lambda1640$ ratio(Figure~\ref{fig:6}b) is an $E_{\rm peak}$ diagnostic that has similar parameter dependence to the H$\beta$/[He~II]$\,\lambda4686$ ratio. 

In general, we recommend H$\beta$/[He~II]$\,\lambda4686$ when [He~II]$\,\lambda4686$ is detected with sufficient signal to noise because these two lines have much smaller separation in wavelength and are therefore less sensitive to dust extinction. However, [He~II]$\,\lambda4686$ is an order of magnitude weaker than He~II$\,\lambda1640$, making H$\beta$/He~II$\,\lambda1640$ an alternative option when [He~II]$\,\lambda4686$ is not sufficiently detected in AGN spectra.

\subsection{$E_{\rm peak}$ Diagnostics with Four Emission Lines}

The other kind of $E_{\rm peak}$ diagnostics are obtained using the ratios of emission lines from the different ionization states of the same atomic species. A harder radiation field contains a larger fraction of high-energy ionizing photons. This hard radiation field leads to a higher fraction of highly ionized species for the same atoms (i.e., argon, neon, sulfur, oxygen), resulting in a larger line ratio of emission line fluxes from high-ionized species to low-ionized species. However, most of these emission line ratios are more sensitive to the ionization parameter, making it challenging to separate the degeneracy between $E_{\rm peak}$ and the ionization parameter using only one emission line ratio.

Fortunately, the emission line ratios from different atoms have different dependences on the $E_{\rm peak}$ and the ionization parameter, making it possible to cancel out the effect from the ionization parameter via a carefully chosen combination of emission line ratios from two different atoms. Here we present one optical $E_{\rm peak}$ diagnostics that fall into this category: 
$$\scalemath{0.8}{\rm R(Ne35,O32,Op)=(\frac{\rm [Ne~III]\,\lambda3869}{\rm [Ne~V]\,\lambda3426})^{0.5}\times(\frac{\rm [O~III]\,\lambda5007}{\rm [O~II]\,\lambda3727})^{1.4}}$$

Using the emission lines [Ne~V]$\,\lambda3426$, [Ne~III]$\,\lambda3869$, [O~II]$\,\lambda\lambda3726,29$, and [O~III]$\,\lambda5007$, R(Ne35, O32, Op) (Figure~\ref{fig:6}d) is an ideal optical $E_{\rm peak}$ diagnostic for moderate to hard AGN radiation fields {\color{black}($\log(E_{\rm peak}/\rm keV)>-1.5$, where R(Ne35,O32,Op)$\lesssim$0.8),} where the metallicity, ionization parameter, and gas pressure only have a small impact on the $E_{\rm peak}$ diagnostic (the R(Ne35,O32,Op) ratio varies up to$\sim0.2\,$dex for $7.3<12+\log(\rm O/H)<9.4$, $-3.4<\log(U)<-2.2$, and $6.2<\log(P/k)<7.8$). The nitrogen abundance has almost no effect on this $E_{\rm peak}$ diagnostic except for at extremely high metallicity ($12+\log(\rm O/H)\approx9.4$), where the nitrogen abundance can also affect the $E_{\rm peak}$ measurement by introducing a $\sim0.1\,$dex variation (see Figure~\ref{fig:B6} in Appendix B). 
 
\begin{deluxetable}{c|c|c|c|c}[t]
\tabletypesize{\scriptsize}
\tablewidth{0pc}
\tablecaption{Cubic Surface Fits for $E_{\rm peak}$ Diagnostics for AGN Narrow Line Regions with $\log(P/k)=7.0$ and `NHlow' Scaling Relation\label{tab:6}}
\tablenum{7}
\tablehead{%\multicolumn{7}{c}{\tiny $F=a_1X+a_2Y+a_3Z+a_4X^2+a_5XY+a_6XZ+a_7Y^2+a_8YZ+a_9Z^2+a_{10}X^3$}\\
%\multicolumn{7}{c}{\tiny $+a_{11}X^2Y+a_{12}X^2Z+a_{13}XY^2+a_{14}XYZ+a_{15}XZ^2+a_{16}Y^3+a_{17}Y^2Z+a_{18}YZ^2$}\\ 
\multicolumn{5}{c}{$X=\log(\rm O/H)+12$, $Y=\log(U)$, $Z=\log(R)$, $F=\log(E_{\rm peak}/\rm keV)$,}\\
\multicolumn{5}{c}{$7.29\leq X\leq9.41$, $-3.8 \leq Y\leq-1.0$}\\
\hline
\colhead{R} & \colhead{$\rm \frac{HeII\,\lambda1640}{HeI\,\lambda3187}$} & \colhead{$\rm \frac{H\beta}{HeII\,\lambda1640}$} & \colhead{$\rm \frac{H\beta}{[HeII]\,\lambda4686}$} & \colhead{R(Ne35,O32,Op)\tablenotemark{\tiny a}}}
\startdata
$F_{\rm min}$ & -2.0 & -2.0 & -2.0 & -1.75 \\
$F_{\rm max}$ & -1.0 & -1.0 & -1.0 & -0.75 \\
\hline
$a_1$ & 16.951 & 8.192 & 9.901 & 32.853 \\
$a_2$ & -2.934 & 0.019 & 1.186 & 7.73 \\
$a_3$ & 7.074 & -1.122 & -5.338 & 0.003 \\
$a_4$ & -1.905 & -0.932 & -1.166 & -4.113 \\
$a_5$ & 0.496 & 0.131 & -0.047 & -0.602 \\
$a_6$ & -1.016 & -0.202 & -0.135 & 0.103 \\
$a_7$ & -0.25 & 0.072 & 0.147 & 1.822 \\
$a_8$ & 1.367 & -0.401 & -0.469 & 0.224 \\
$a_9$ & -0.989 & 1.772 & 3.479 & -0.408 \\
$a_{10}$ & 0.071 & 0.033 & 0.041 & 0.165 \\
$a_{11}$ & -0.036 & -0.019 & -0.012 & -0.018 \\
$a_{12}$ & 0.04 & 0.027 & 0.051 & -0.023 \\
$a_{13}$ & -0.022 & -0.023 & -0.024 & -0.147 \\
$a_{14}$ & -0.062 & 0.018 & 0.051 & -0.032 \\
$a_{15}$ & 0.078 & -0.167 & -0.199 & 0.064 \\
$a_{16}$ & -0.048 & -0.025 & -0.024 & 0.076 \\
$a_{17}$ & 0.135 & -0.082 & -0.074 & -0.038 \\
$a_{18}$ & -0.152 & -0.159 & -0.12 & -0.016 \\
$a_{19}$ & 0.073 & -0.483 & -0.514 & -0.044 \\
$a_{20}$ & -53.223 & -24.512 & -25.551 & -82.729 \\
\hline
RMS err & 4.15\% & 3.43\% & 3.76\% & 4.87\% \\
%R2_1 & 0.9859 & 0.9904 & 0.9884 & 0.9729 \\
%R2_2 & 0.8813 & 0.9019 & 0.8923 & 0.8353 \\
%score & 0.9859 & 0.9904 & 0.9884 & 0.9729 \\
\enddata
\tablecomments{Tricubic surface fit:$Z=a_1X+a_2Y+a_3Z+a_4X^2+a_5XY+a_6XZ+a_7Y^2+a_8YZ+a_9Z^2+a_{10}X^3+a_{11}X^2Y+a_{12}X^2Z+a_{13}XY^2+a_{14}XYZ+a_{15}XZ^2+a_{16}Y^3+a_{17}Y^2Z+a_{18}YZ^2+a_{19}Z^3+a_{20}$}
\tablenotetext{\tiny a}{R(Ne35,O32,Op)=$(\frac{\rm [Ne~III]\,\lambda3869}{\rm [Ne~V]\,\lambda3426})^{0.5}\times(\frac{\rm [O~III]\,\lambda5007}{\rm [O~II]\,\lambda3727})^{1.4}$}
\end{deluxetable}
\section{Diagnostic testing}
	\label{sec:comp}
	
{\color{black} The goal of this work is to provide metallicity diagnostics for AGN spectra. The ionization parameter, $E_{\rm peak}$, and gas pressure diagnostics are provided as supplemental diagnostics to help investigators account for secondary dependencies on these parameters.}

In metallicity studies on large samples of galaxies, it is crucial to obtain reliable and consistent metallicity measurements for all galaxies in the sample, which could be affected by the choice of metallicity diagnostics and the mixing of Seyfert galaxies and pure star-forming galaxies. 

In this section, we show that the AGN diagnostics presented in this paper are ideal for metallicity studies on a large sample of galaxies through two tests: (i) One external test showing that our diagnostics are consistent with AGN diagnostics presented in the literature. (ii) One observational test showing that different metallicity diagnostics in this paper provide consistent metallicity measurements on the same sample of galaxies with the same optical and UV spectra.

\subsection{External Testing: Comparison with Literature}

\begin{figure*}[htb]
\epsscale{1.1}
\plotone{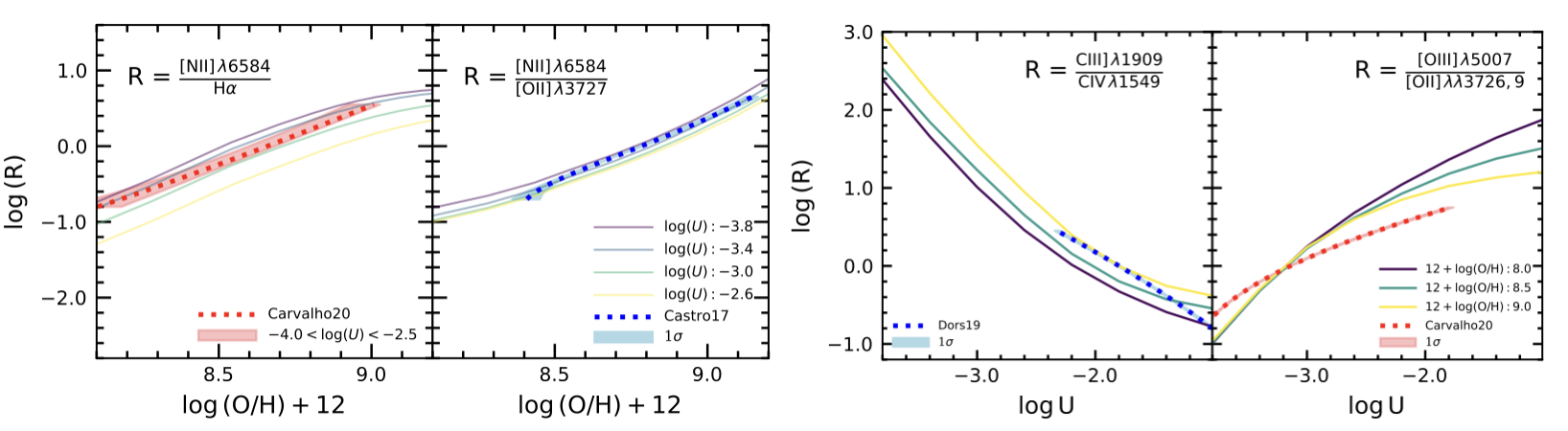}
\caption{External testing the AGN metallicity diagnostics and ionization parameter diagnostics in this paper. The left panel shows the comparison of optical AGN metallicity diagnostics [N~II]/H$\alpha$ and [N~II]/[O~II] from this paper and those from \citet{carvalho_chemical_2020} and \citet{castro_new_2017}. The right panel shows comparison of AGN ionization parameter diagnostic C~III]/C~IV and [O~III]/[O~II] from this paper and those from \citet{dors_semi-empirical_2019} and \citet{carvalho_chemical_2020}. In both panels, the diagnostics from this paper are derived from AGN models with $\log(P/k)=7.0$, $\log(E_{\rm peak}/\rm keV)=-1.5$, and `NHhigh' nitrogen abundance scaling relation.
\label{fig:7}}
\end{figure*}

There are several UV and optical AGN diagnostics available in the literature (e.g.,\citet{storchi-bergmann_chemical_1998,nagao_gas_2006,dors_central_2015,castro_new_2017,carvalho_chemical_2020}). For the external testing, we select four recent and well-tested diagnostics from the literature: (1) metallicity diagnostics [N~II]$\,\lambda6584$/[O~II]$\,\lambda\lambda3727,9$ from \citet{castro_new_2017}; (2) [N~II]$\,\lambda6584$/H$\alpha$ from \citet{carvalho_chemical_2020}; (3) the ionization parameter diagnostics C~III]$\,\lambda1909$/C~IV$\,\lambda1549$ from \citet{dors_semi-empirical_2019}; (4) the optical [O~III]$\,\lambda5007$/[O~II]$\,\lambda\lambda3727,9$ ionization parameter diagnostics from \citet{carvalho_chemical_2020}. These external diagnostics are derived using a dust-free AGN model with the 'NHhigh' nitrogen scaling relation.

We compare the above diagnostics derived from our AGN model with $\log(E_{\rm peak}/\rm keV)=-1.5$ and the 'NHhigh' nitrogen scaling relation to the external diagnostics. As shown in Figure~\ref{fig:7}, the metallicity diagnostics [N~II]/H$\alpha$ and [N~II]/[O~II] derived from our AGN model are consistent with the diagnostics in \citet{castro_new_2017} and \citet{carvalho_chemical_2020}. For the UV ionization parameter diagnostics C~III]/C~IV, \citet{dors_semi-empirical_2019} accounts for the variation of metallicity in a sample of Seyfert galaxies and obtain the best-fit relation between the ratio of C~III]/C~IV and the ionization parameter. This best-fit relation is consistent with our theoretical diagnostic with metallicity spanning the range in $8.0\lesssim12+\log(\rm O/H)\lesssim9.0$.

The two optical [O~III]/[O~II] diagnostics have similar trends with the ionization parameter, although they are offset by $\sim0.4\,$dex in the line ratio in the comparison. This difference is a result of different dust assumptions in the two models. Our AGN model includes dust and dust depletion, while the AGN model in \citet{carvalho_chemical_2020} is dust-free. The inclusion of dust introduces photoelectric heating that leads to a higher temperature in the nebula, resulting in the higher [O~III]/[O~II] ratio at high ionization parameter $-3.0<\log(U)$ and the lower [O~III]/[O~II] ratio at low ionization parameter $\log(U)<-3.0$ compared with to the dust-free diagnostic.

To summarize, despite the apparent difference in [O~III]/[O~II] diagnostic caused by different dust settings, {\color{black} our theoretical diagnostics are generally consistent with the literature when our AGN models use the same nitrogen abundance scaling relation (`NHhigh').}

Nevertheless, as mentioned in previous sections and \citet{zhu_new_2023}, nitrogen abundance and AGN ionizing spectra can vary in different Seyfert galaxies. The ability of our AGN model to account for the intrinsic variation in nitrogen abundance and AGN ionizing spectrum in Seyfert galaxies provides a broader application range for the theoretical diagnostics presented in this paper.

\subsection{Observational Testing}	

\begin{figure}[htb]
\epsscale{1.3}
\plotone{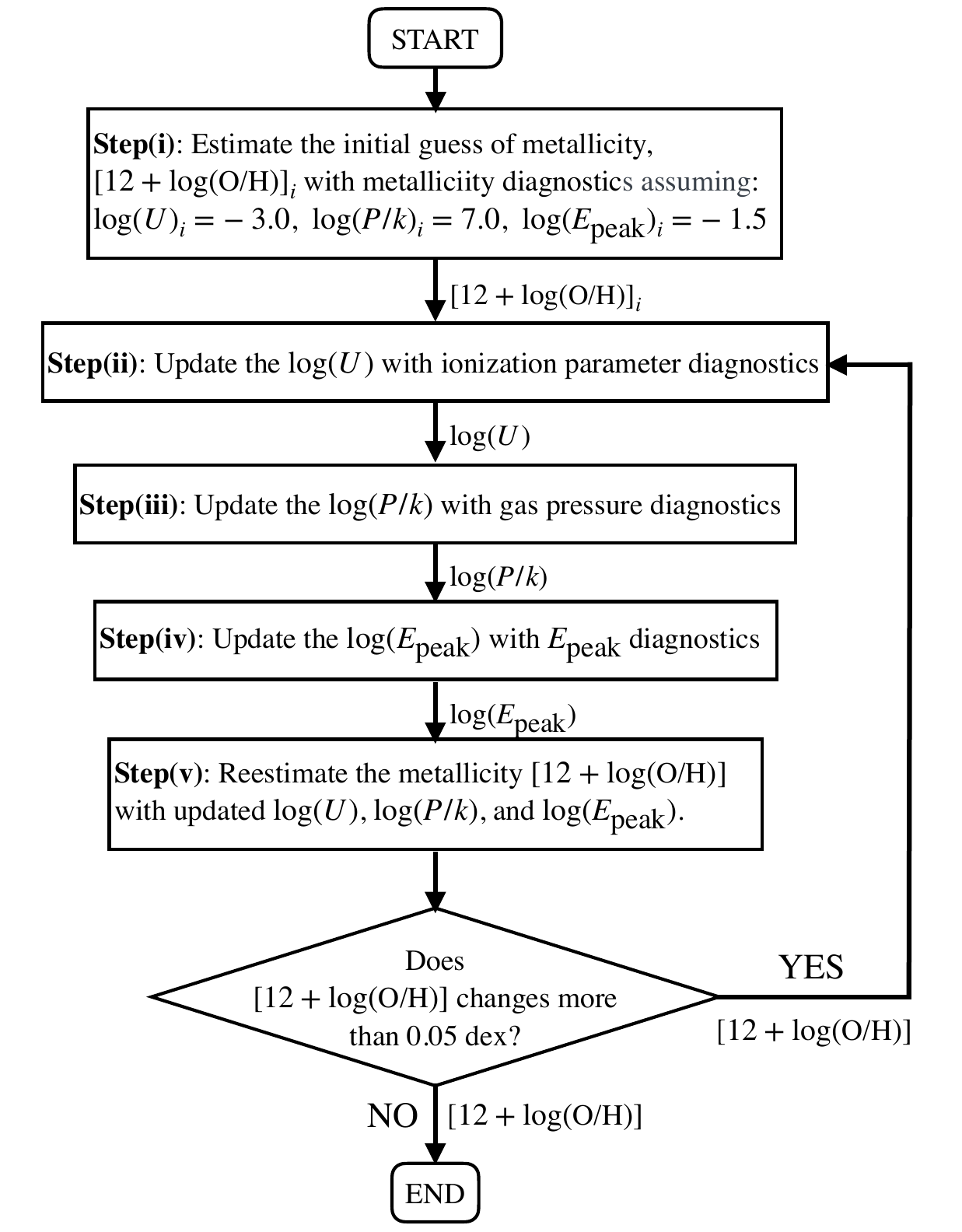}
\caption{{\color{black}The logical flowchart for determining metallicity for AGN narrow line regions using theoretical diagnostics.}
\label{fig:20}}
\end{figure}

\begin{figure*}[htb]
\epsscale{1.1}
\plotone{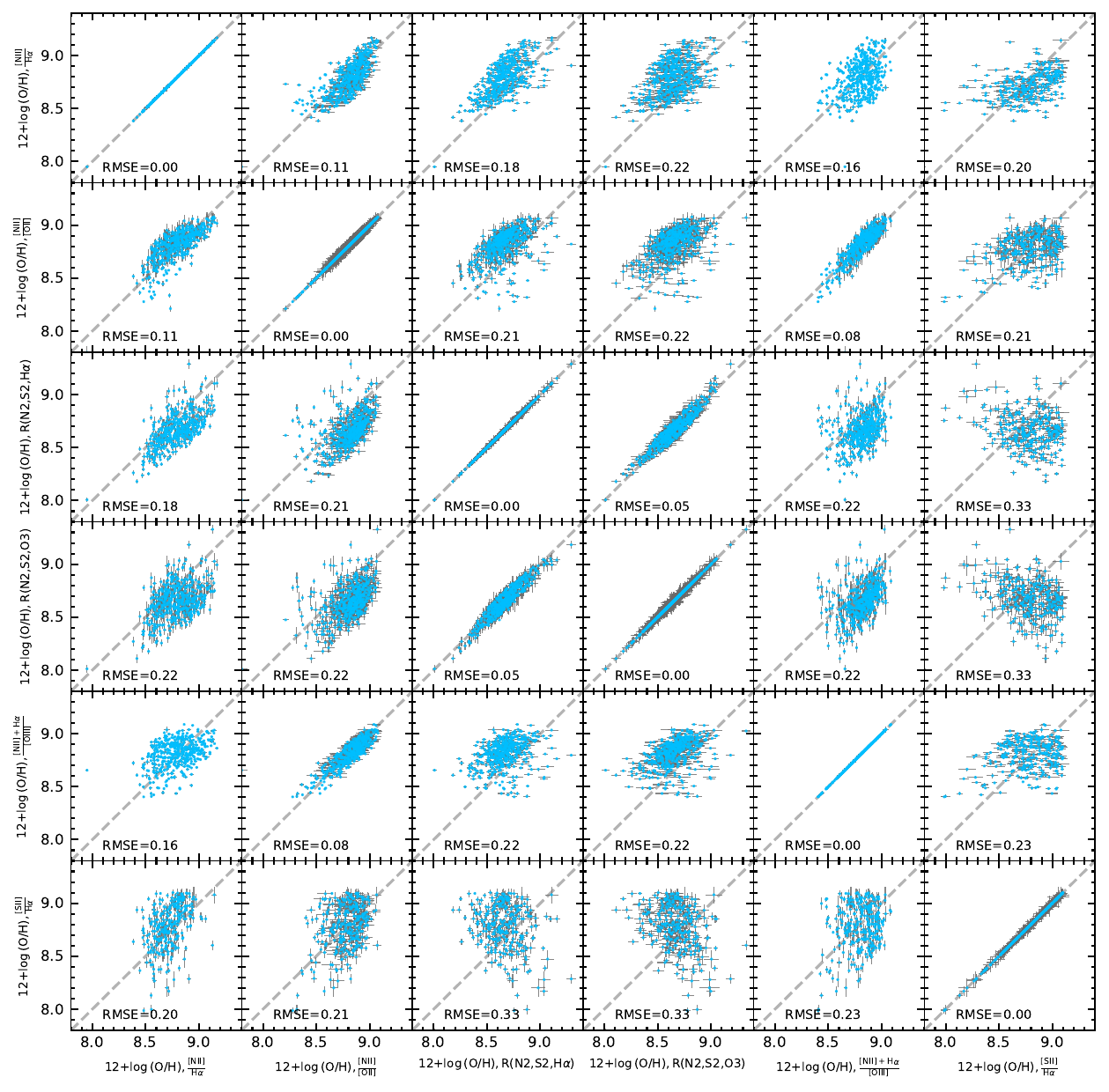}
\caption{Comparing the metallicity measurements from six optical AGN metallicity diagnostics using `NHhigh' nitrogen scaling relation for 460 Seyfert galaxies observed in SDSS DR16. From left to right (from top to bottom), the x-axis (y-axis) represents the metallicity measurements from [NII]/H$\alpha$, [NII]/[OII], R(N2, S2, H$\alpha$), R(N2, S2, O3), ([NII]+H$\alpha$)/[OIII], and [SII]/H$\alpha$. The metallicity measurements on SDSS spectra are shown in blue dots, with error bars indicating the measurement errors. The Root Mean Square Errors (RMSE) are shown in every subplot.
\label{fig:8}}
\end{figure*}

To perform the observational test, we need a sample of Seyfert galaxies with a wide wavelength coverage in their spectra to contain sufficient emission lines for cross-matching different diagnostics. 

At optical wavelengths, such samples can be obtained through the Sloan Digital Sky Survey (SDSS). We select our sample from SDSS data release 16 (DR16; \citet{kollmeier_sdss-v_2017}). Following the classification method in \citet{zhu_new_2023} and applying a Signal-to-noise ratio (S/N)$>5$ limit to the selected emission lines, we select a sample of 460 Seyfert galaxies that contain [O~II]$\,\lambda\lambda3727,9$, [O~III]$\,\lambda4959$, [O~III]$\,\lambda5007$,  [N~II]$\,\lambda6584$, [S~II]$\,\lambda\lambda6717,31$, H$\alpha$, and H$\beta$. 

{\color{black} Prior to the testing, we performed extinction correction for these optical emission-line fluxes with the Balmer decrement and applied the relative extinction curve with $R_V^A=4.5$ and $A_V=1$ of \citet{2005ApJ...619..340F}. We assume an intrinsic H$\alpha$/H$\beta$=3.1 for these Seyfert galaxies \citep{vogt_galaxy_2013}.}

Using the SDSS spectra, we measure the metallicity of these Seyfert galaxies using the diagnostics [N~II]/H$\alpha$, [N~II]/[O~II], R(N2,S2,H$\alpha$), R(N2,S2,O3), ([N~II]+H$\alpha$)/[O~III], and [S~II]/H$\alpha$ in five steps {\color{black}(as shown in Figure~\ref{fig:20})}:

{\color{black}Step(i). Estimate the initial guess of metallicity. In this step, we assume initial gas pressure $\log(P/k)=7.0$, AGN radiation field $\log(E_{\rm peak}/\rm keV)=-1.5$, and ionization parameter $\log(U)=-3.0$ for all galaxies and use the above metallicity diagnostics to estimate the metallicities for all galaxies using the cubic fits. 

Step(ii). Update the ionization parameter using the [O~III]$\,\lambda5007$/[O~II]$\,\lambda3727,9$ ratio. In this step, the metallicity estimated in Step(i) will be used in the cubic fits. 
%We then obtain the mean gas pressure $\log(P/k)=7.0$ for all galaxies and apply this mean value in the following calculations, which is a reasonable proxy since all diagnostics used in the following steps are insensitive to gas pressure.

Step(iii). Update gas pressure using the [O~II]$\,\lambda3729$/[O~II]$\,\lambda3727$ ratio. In this step, the metallicity estimated in Step(i) and the ionization parameter updated in Step(ii) will be used in the cubic fits.

Step(iv). Update the $E_{\rm peak}$ using the [He~II]$\,\lambda4686$/H$\beta$ ratio. In this step, the metallicity estimated in Step(i), the updated ionization parameter, and gas pressure in Step(ii) and (iii) will be used in the cubic fit. In most galaxies, the measured [He~II]$\,\lambda4686$ fluxes do not satisfy the S/N$>3$ requirement. The mean $E_{\rm peak}$ values from galaxies that detect [He~II]$\,\lambda4686$ with S/N$>3$ is $\log(E_{\rm peak}/\rm keV)=-1.68$. This mean value will be applied to all Seyfert galaxies in the sample in the following calculations.

Step(v). Reestimate the metallicity using the updated ionization parameter $\log(U)$, gas pressure $\log(P/k)$, and $E_{\rm peak}$. If the difference between the re-estimate metallicity and previous metallicity is larger than 0.05\,dex, we repeat Step(ii) to Step (v). When the difference becomes smaller than 0.05\,dex, we adopt the re-estimated metallicity as the final value. The measurement errors are calculated from the errors of emission line fluxes and the root mean square errors of the cubic fits. For these 460 Seyfert galaxies, the iteration times to reach convergence on the metallicity measurement are typically two to four.}
%The metallicities of these galaxies exceed the valid measurement range of the metallicity diagnostics ([O~III]+[O~II])/H$\beta$, which are thus not included in this comparison.

In Figure~\ref{fig:8}, we present the cross-comparison of the metallicity measurements using diagnostics derived from AGN models with the `NHhigh' nitrogen scaling relation. These six optical metallicity diagnostics are generally consistent with each other, with the root mean square errors (RMSE) $\lesssim0.33$\,dex. The [N~II]/H$\alpha$ and [N~II]/[O~II] metallicities present the best agreement with all other diagnostics, showing the least RMSEs among the six diagnostics. The similar test on diagnostics derived from AGN models using `NHlow' nitrogen scaling relation presents similar consistency, only that the measured metallicities are systematically higher by $\sim0.3\,$dex. This is expected since all six metallicity diagnostics have a strong dependence on the nitrogen scaling relation, as shown in Figure~\ref{fig:B2}.

{\color{black}On the other hand, the [S~II]/H$\alpha$ metalicity presents the largest scatter with the other five diagnostics, suggesting it is not ideal for metallicity measurement in this sample. In addition to the measurement errors, the scatters in the comparison are also caused by the simple assumption of uniform $E_{\rm peak}$ for all Seyfert galaxies and the mixing of HII regions and AGN NLRs in the field of view of the SDSS spectra. The mixture of HII regions in AGN-dominated spectra can underestimate the metallicity derived from [S~II]/H$\alpha$ and [N~II]/H$\alpha$ ratios and overestimate the metallicities derived from all other four diagnostics (See discussion in Section~\ref{sec:8}). 

The $E_{\rm peak}$ measurement can be improved for galaxies in our sample with more complete detections of [He~II] $\,\lambda4686$ or [Ne~V]$\,\lambda3426$. The mixture of HII regions in AGN-dominated can be disentangled with spatially resolved spectra (i.e., integral field spectroscopy).}
%As shown in Section~\ref{sec:8}, emission line ratios generated from HII regions have different relations to metallicity, which can mislead the metallicity measurements in Seyfert galaxies if spectra are assumed to be pure-AGN when HII regions contaminate them. At the same value of some diagnostics, the metallicity in pure AGN regions can be up to 1.0 dex higher than in pure HII regions.

\begin{figure}[htb]
\epsscale{1.3}
\plotone{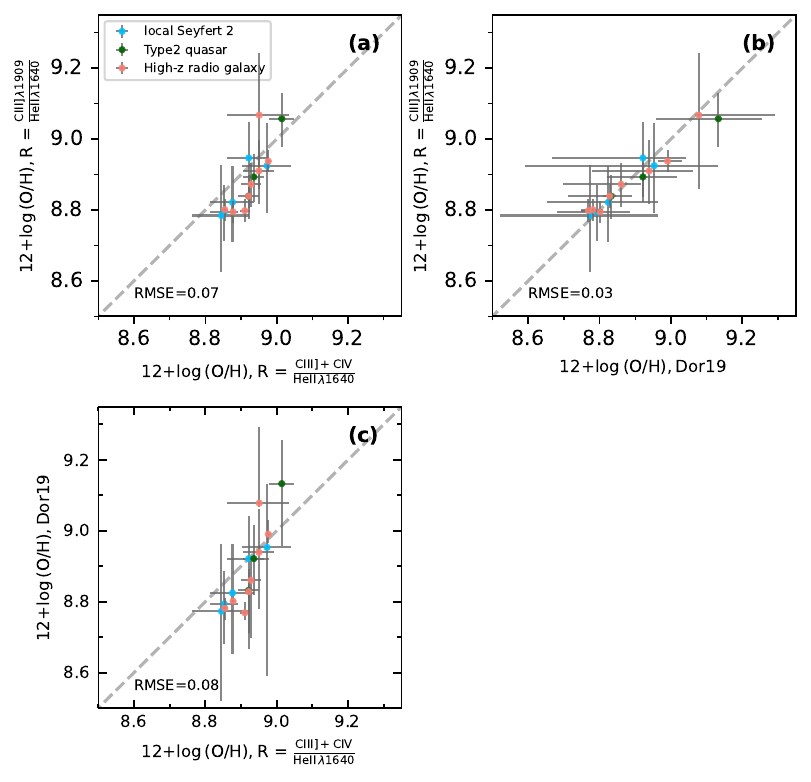}
\caption{{\color{black} Comparing the metallicity measurements between (a)the two UV AGN metallicity diagnostics C~III]/He~II and C$_{34}$/He~II presented in this paper, (b)the C~III]/He~II metallicity and the metallicity measured in \citet{dors_semi-empirical_2019}, and (c)the C$_{34}$/He~II metallicity and the metallicity measured in \citet{dors_semi-empirical_2019}. }
\label{fig:9}}
\end{figure}

Following the similar metallicity measurement approach as in optical wavelength, we test the performance of two UV metallicity diagnostics C~III]/He~II and C$_{34}$/He~II on a sample of {\color{black} Type-2 Seyfert galaxies at redshift $0<z<4.0$ adopted from \citet{dors_semi-empirical_2019}, who assemble the data from \citet{kraemer_iue_1994,de_breuck_sample_2000,nagao_gas_2006,bornancini_imaging_2007,matsuoka_chemical_2009,matsuoka_mass-metallicity_2018}. \citet{dors_semi-empirical_2019} also estimated the metallicity for these galaxies using the combination of C$_{34}$/He~II and C~III]/C~IV ratios. Among the 77 Seyfert galaxies in their original sample, only 28 galaxies have measurement errors for their emission line ratios. Given the large measurement errors in these galaxies, we only use the 28 galaxies that contain measurement errors in their emission line ratios in our test for analysis.

 Due to the limited wavelength coverage of the spectra, gas pressure diagnostics and $E_{\rm peak}$ diagnostics are not possible for this sample. Therefore, we estimate the metallicity with only Steps (i), (ii), and (v), but assuming a uniform gas pressure $\log(P/k)=7.0$ and a uniform AGN radiation field $\log(E_{\rm peak}/\rm keV)=-1.0$. We estimate the ionization parameter using the C~III]$\,\lambda1909$/C~IV$\,\lambda1549$ ratio. To solve the metallicity degeneracy in the C~III]/He~II and C$_{34}$/He~II diagnostics, we refer to the metallicity measurement in \citet{dors_semi-empirical_2019} to determine the branch used in our measurement.

%Among the 77 Seyfert galaxies, only 39 galaxies are within the metallicity measurement range for C~III]/He~II (C~III]/He~II$\lesssim-$0.4) and C$_{34}$/He~II (where C$_{34}$/He~II$\lesssim$0.1). As mentioned in Section~\ref{sec:metal}, both C~III]/He~II and C$_{34}$/He~II are double-valued to metallicity. We use the metallicity in \citet{dors_semi-empirical_2019} to determine the branch used in our measurement.

Figure~\ref{fig:9} presents the metallicity measurements for 16 galaxies whose line ratios are within the valid ranges for C~III]/He~II and C$_{34}$/He~II metallicity diagnostics. As shown in Figure~\ref{fig:9}, the metallicity measurements from the two UV metallicity diagnostics are consistent with each other to within $\sim0.2$\,dex. The comparisons between our metallicity measurements and the metallicity measured in \citet{dors_semi-empirical_2019} are consistent (RMSE$\lesssim$0.1\,dex). However, the large errors from the emission line ratios lead to significant errors in all three metallicity measurements. A larger sample of Seyfert galaxies with higher S/N UV observations is needed to provide a comprehensive test of our UV diagnostics for AGN.}

\section{Caveat: The Bias of Metallicity Measurement in HII and AGN Mixed Galaxies\label{sec:8}}

\begin{figure*}[htb]
\epsscale{1.1}
\plotone{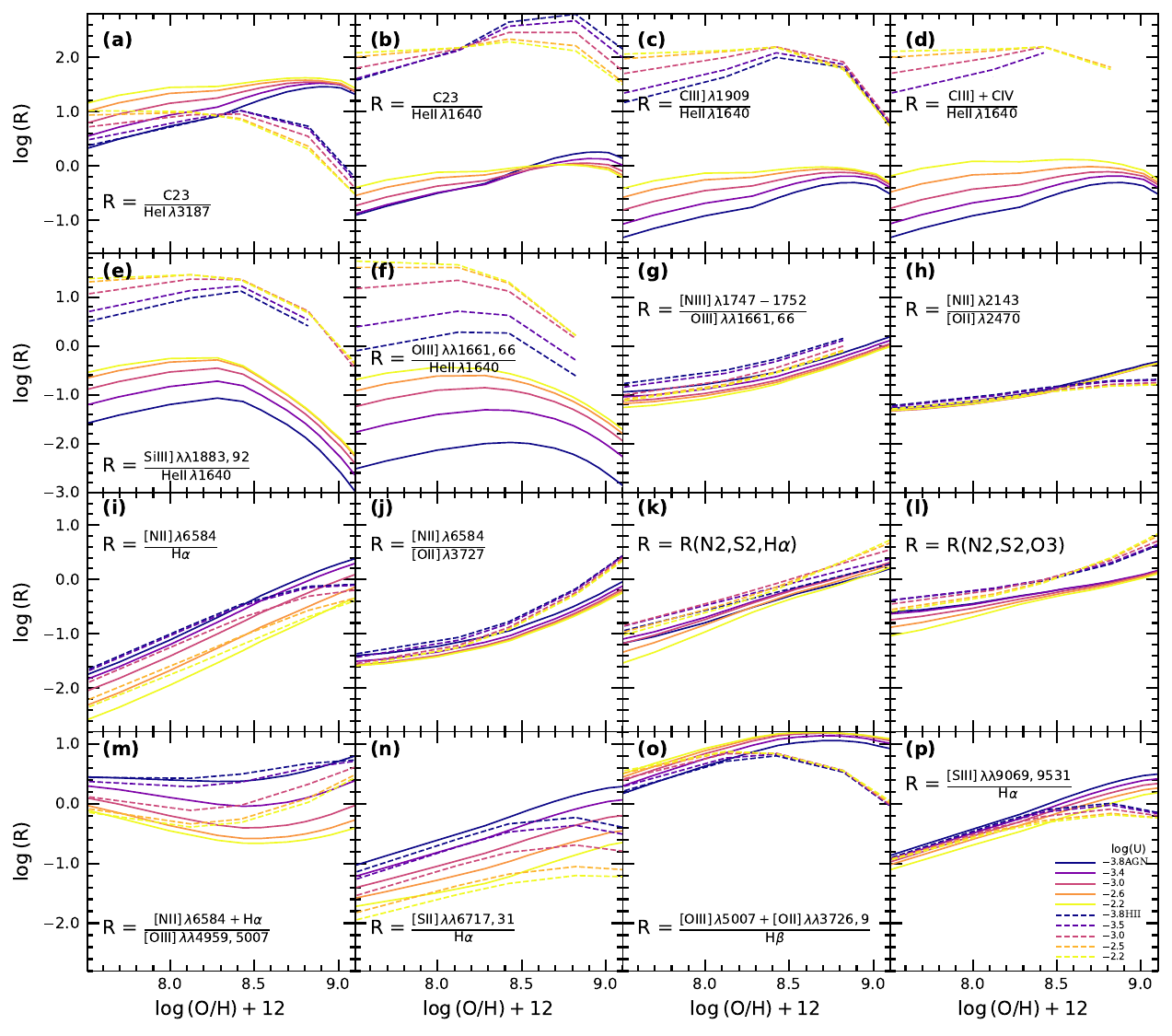}
\caption{{\color{black}Comparing the UV and optical metallicity diagnostics for AGN narrow line regions (solid lines) and HII region (dashed lines) with varied ionization parameter ($\log(U)=-3.8,-3.4,-3.0,-2.6,-2.2$ correspond to colors from dark to bright). Both AGN and HII metallicity diagnostics are derived from models with the same gas pressure $\log(P/k)=7.0$.}
\label{fig:11}}
\end{figure*}

Before theoretical diagnostics in the AGN region became available, it was common to assume that the theoretical diagnostics derived for HII regions were approximately the same in AGN regions and thus used in metallicity measurement in the AGN regions. However, this simple assumption can lead to significant errors for most metallicity diagnostics, especially at the high metallicity regime. 

%In addition, due to the difficulty in disentangling HII-AGN mixing in single-aperture galaxy spectra, the metallicity measurement for composite galaxies can be changed by up to $\sim1$\,dex when switching from HII metallicity diagnostics to AGN metallicity diagnostics.
%
{\color{black} In this section, we explore the differences in theoretical diagnostics for the gas metallicity, ionization parameter, and gas pressure between HII region models and AGN models to test whether the HII region diagnostics could be applied to AGN narrow-line region gas, as done in the past. 

To perform the investigation, we adopt the HII region model from \citet{kewley_theoretical_2019} and our AGN model with `NHlow' nitrogen scaling relation from \citet{zhu_new_2023}. In both the HII region model and AGN model, MAPPINGS version 5.2 \citep{sutherland_mappings_2018}, CHIANTI version 10 database \citep{del_zanna_chiantiatomic_2021}, the abundance scaling relations from \citet{2017MNRAS.466.4403N}, and dust depletion factors from \citet{jenkins_unified_2009,jenkins_depletions_2014} are used in the model calculation. This guarantees consistent physics and atomic data in our comparison of theoretical diagnostics. 

The stellar ionizing radiation field in the HII region model is generated from Starburst99. Detail description of the Starburst99 model can be found in \citet{levesque_theoretical_2010,nicholls_resolving_2012,kewley_theoretical_2019}. Briefly, this model adopts the Salpeter initial mass function with an upper mass limit of 100$M_\odot$ \citep{salpeter_luminosity_1955}, uses the Pauldrach/Hillier model for stellar atmosphere \citep{hillier_treatment_1998,pauldrach_radiation-driven_2001}, and the Geneva group ``high'' mass-loss evolutionary tracks \citep{meynet_grids_1994}. The stellar ionizing radiation field is then obtained at the stellar age of 5\,Myr with a continuous star formation history. For the AGN model, we use the $\log(E_{\rm peak}/\rm keV)=-1.5$ AGN ionizing radiation field in this comparison.

%update the text accordingly.
In Figure~\ref{fig:11}, we compare the theoretical HII and AGN metallicity diagnostics at UV and optical wavelengths. As shown in Figure~\ref{fig:11}, the most consistent HII and AGN metallicity diagnostics are [N~II]$\,\lambda2143$/[O~II]$\,\lambda2470$, [N~II]$\,\lambda6584$/H$\alpha$, [N~II]$\,\lambda6584$/[O~II]$\,\lambda\lambda3727,9$, and [S~III]$\,\lambda\lambda9069,9531$/H$\alpha$. At $12+\log(\rm O/H)\lesssim8.5$, the line ratios of these four diagnostics vary less than 0.1\,dex between the AGN and HII models at the same metallicity. 

In the high metallicity regime ($12+\log(\rm O/H)\gtrsim8.5$), almost all HII and AGN metallicity diagnostics largely different. For example, with a 50\% contribution in flux from HII regions, the metallicity measured with AGN metallicity diagnostics [N~II]$\,\lambda2143$/[O~II]$\,\lambda2470$ and [N~II]$\,\lambda6584$/H$\alpha$ can be underestimated by $\sim0.2$\,dex. In contrast, when using AGN metallicity diagnostic [N~II]$\,\lambda6584$/[O~II]$\,\lambda\lambda3727,9$, the metallicity measurement can be overestimated by $\sim0.1$\,dex.

\begin{figure*}[htb]
\epsscale{1.1}
\plotone{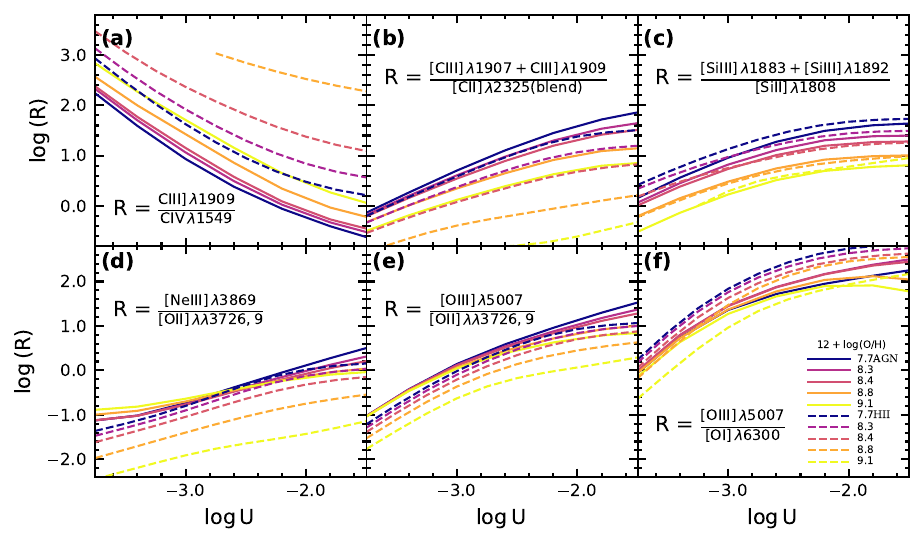}
\caption{{\color{black}Comparing the UV and optical ionization parameter diagnostics for AGN narrow line regions (solid lines) and HII region (dashed lines) with varied metallicity ($12+\log(\rm O/H)=7.7,8.3,8.4,8.8,9.1$ correspond to colors from dark to bright). Both AGN and HII ionization parameter diagnostics are derived from models with the same gas pressure $\log(P/k)=7.0$.}
\label{fig:12}}
\end{figure*}

%[N~III]$57\rm\mu m$/[O~III]$52\rm\mu m$ at the far-IR wavelength, with $\Delta\log(R)\lesssim0.2$ in the full range of metallicity. The consistent relationship between [N~III]$57\rm\mu m$/[O~III]$52\rm\mu m$ and metallicity across AGN and HII regions and the fact that this relationship is insensitive to the ionization parameter and $E_{peak}$ makes [N~III]$57\rm\mu m$/[O~III]$52\rm\mu m$ an ideal metallicity diagnostic for metallicity measurement in Seyfert galaxies.

The line ratios $C_{23}$/He~I, $C_{23}$/He~II, C~III]/He~II, $C_{34}$/He~II, Si~III]/He~II, and O~III]/He~II in UV wavelength exhibit a large discrepancy (line ratios change by $\gtrsim1.0$\,dex between the HII and AGN diagnostics), across the full range of metallicity. Applying these six HII metallicity diagnostics to a spectrum containing an AGN will lead to significant measurement errors.}

For the remainder of the metallicity diagnostics, we observed consistent line ratios between AGN and HII models at $12+\log(\rm O/H)\lesssim8.0$, with line ratios varying $\lesssim0.1$ between AGN and HII regions at the same metallicity. This indicates that these HII metallicity diagnostics can approximate the metallicity in AGN regions at low metallicity. However, as the metallicity increases and exceeds $12+\log(\rm O/H)\approx8.0$, the AGN metallicity diagnostics deviate from the HII metallicity diagnostics. The deviation in line ratios between AGN and HII regions can increase up to 1.0\,dex at the same metallicity.

\begin{figure*}[htb]
\epsscale{1.1}
\plotone{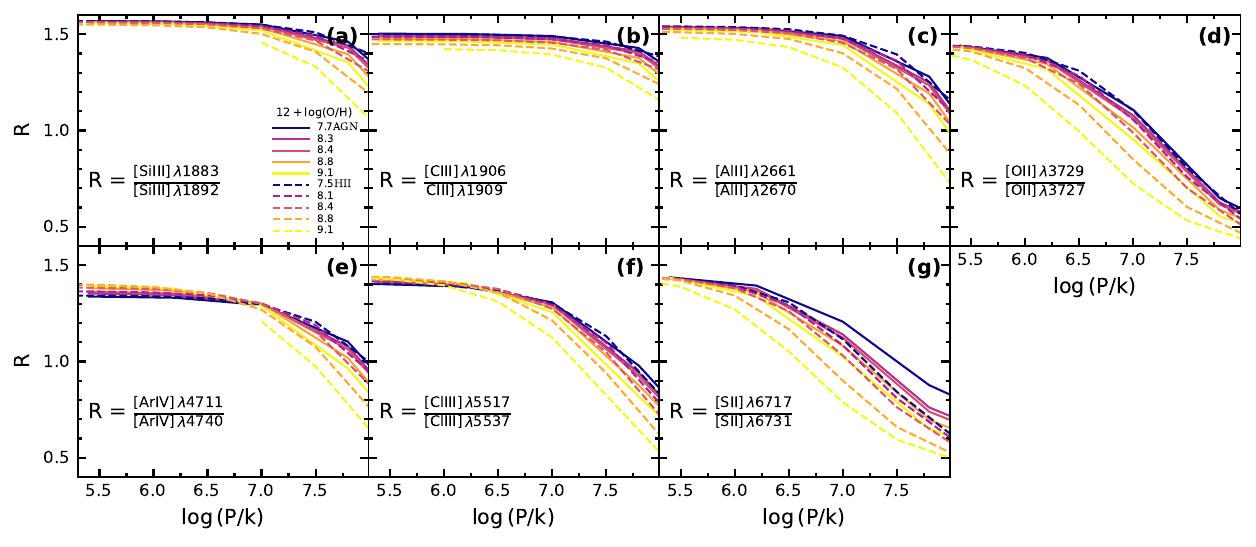}
\caption{{\color{black}Comparing the UV and optical gas pressure diagnostics for AGN narrow line regions (solid lines) and HII region (dashed lines) with varied metallicity ($12+\log(\rm O/H)=7.7,8.3,8.4,8.8,9.1$ correspond to colors from dark to bright). Both AGN and HII metallicity diagnostics are derived from models with the same ionization parameter $\log(U)=-3.0$.}
\label{fig:13}}
\end{figure*}

{\color{black}As shown in the top panel of Figure~\ref{fig:12}, large discrepancies between the AGN and HII models are also found in ionization parameter diagnostics, especially at the high metallicity regime ($12+\log(\rm O/H)\gtrsim8.4$). Among all, [Si~III]$\,\lambda\lambda1883,92$/[Si~II]$\,\lambda1808$ is the most consistent HII and AGN UV ionization parameter diagnostic. In the optical, [Ne~III]$\,\lambda3869$/[Ne~V]$\,\lambda3426$ and [O~III]/[O~II] produce similar metallicities at $12+\log(\rm O/H)\gtrsim8.4$. Nevertheless, at $12+\log(\rm O/H)\gtrsim8.4$ or for other ionization parameter diagnostics, the line ratios can differ up to $\sim1.0$\,dex between AGN and HII models.

In contrast to other diagnostics, gas pressure diagnostics are more consistent across AGN and HII regions. This is demonstrated by the Figure~\ref{fig:13}. At $12+\log(\rm O/H)\gtrsim8.4$, most UV and optical gas pressure diagnostics are consistent across AGN narrow line regions and HII regions, except for [S~II]$\,\lambda6717$ and [S~II]$\,\lambda6731$ with line ratio varies up to $\sim0.2$ between AGN region and HII region at the same gas pressure. At the high metallicity regime ($12+\log(\rm O/H)\gtrsim8.4$), AGN gas pressure diagnostics can underestimate the gas pressure by up to $\sim0.25$\,dex with 50\% contribution from HII regions in the flux of the AGN spectra.

This analysis demonstrates that where AGN is expected to dominate the emission-line ratios, AGN diagnostics must be used to estimate the metallicity.  Where HII regions dominate the emission-line ratios, HII regions diagnostics must be used.  HII and AGN diagnostics should not be used interchangeably.  If HII region contamination is expected, we recommend the use of AGN diagnostics which show the greatest consistency with the HII region diagnostics, as described above.}

\section{Conclusion}

{\color{black}This paper presents a set of theoretical diagnostics for gas metallicity, ionization parameter, gas pressure, and the peak energy in AGN ionizing radiation field $E_{\rm peak}$ for AGN narrow-line regions spanning the UV and optical wavelengths. These diagnostics are derived from the new dust-depleted radiation-pressure-dominated AGN model proposed in \citet{zhu_new_2023}. Calculated with MAPPINGS V photoionization code, this new AGN model provides the best predictions for AGN observations across a broad range of wavelengths, including UV, optical, and infrared. 

Using the new AGN model, we investigate the metallicity-sensitive line ratios and study how they vary with metallicity. We also study the dependency of these metallicity diagnostics on the ionization parameter, gas pressure, $E_{peak}$, and the nitrogen scaling relation. We also investigate the ionization parameter diagnostics, gas pressure diagnostics, and $E_{peak}$ diagnostics to help account for the secondary degeneracies in metallicity diagnostics.}

We recommend the following metallicity diagnostics that are most robust against other parameters:

(a) N~III]$\,\lambda$1747-52/O~III]$\,\lambda\lambda1660,6$, only depend on the nitrogen scaling relation.

(b) [N~II]$\,\lambda6584$/[O~II]$\,\lambda\lambda3727,9$, R(N2,S2,H$\alpha$), and R(N2, S2, O3), only depend on the nitrogen scaling relation.

%(c) [Ar~III]$8.99\rm\mu m$/Pf-$\alpha$, [Ne~III]$15.6\rm\mu m$/Br$\alpha$, and ([Ne~III]$15.6\rm\mu m$+[Ne~II]$12.8\rm\mu m$)/Pf$\alpha$, only slightly depend on the $E_{peak}$.

When the above metallicity diagnostics are not available in the observations, other metallicity diagnostics presented in this paper can be used in combination with the ionization parameter diagnostic, gas pressure diagnostics, and $E_{peak}$ diagnostics to obtain a more reliable metallicity for AGNs.

{\color{black}We compare consistent HII and AGN diagnostics and demonstrate that HII and AGN diagnostics should not be used interchangeably.  We recommend the use of AGN diagnostics where AGN is expected to dominate the emission-line gas.  We provide a description of which diagnostics may perform best in the event of contamination by HII gas in the aperture. The metallicity diagnostics that are least affected by the HII-AGN mixing are: }

(i) UV diagnostic [N~II]$\,\lambda\lambda2139,43$/[O~II]$\,\lambda2470$.

(ii) optical diagnostics [N~II]$\,\lambda6584$/H$\alpha$, [N~II]$\,\lambda6584$/[O~II]$\,\lambda\lambda3727,9$, and [S~III]$\,\lambda\lambda9069,9531$/H$\alpha$ at $12+\log(\rm O/H)\lesssim8.5$. 

%(ii) Far-IR diagnostics [N~III]$57\rm\mu m$/[O~III]$52\rm\mu m$ and [N~III]$57\rm\mu m$/[O~III]$88\rm\mu m$ at $12+\log(\rm O/H)\lesssim9.0$.

Through testing, we show that the diagnostics presented in this paper are consistent with those in the literature when using similar parameters in the AGN model. 

UV observations of a large sample of galaxies with wide wavelength coverages are needed to perform a more complete test of our UV diagnostics.

%When available, these metallicity diagnostics can provide a test to the metallicity measured by other diagnostics to identify the contamination of HII-AGN mixing in the observations. When none of these HII-AGN mixing independent diagnostics are available, one should remember that the metallicity measurement might be biased due to the HII-AGN mixing. 

%With a wide wavelength coverage, these AGN diagnostics will enable accurate metallicity studies on both nearby and high-redshift universes, particularly on a large sample of Seyfert galaxies. This has been a missing piece in galaxy evolution studies for a long time. When combined with the HII-AGN separation tool, these diagnostics can provide detailed metallicity maps for a single AGN-host galaxy, which might provide unique constraints on the AGN feedback mechanisms.

\section{Acknowledgement}

Parts of this research were conducted by the Australian Research Council Centre of Excellence for All Sky Astrophysics in 3 Dimensions (ASTRO 3D), through project number CE170100013.

Funding for the Sloan Digital Sky Survey V has been provided by the Alfred P. Sloan Foundation, the Heising-Simons Foundation, the National Science Foundation, and the Participating Institutions. SDSS acknowledges support and resources from the Center for High-Performance Computing at the University of Utah. The SDSS website is www.sdss5.org.

SDSS is managed by the Astrophysical Research Consortium for the Participating Institutions of the SDSS Collaboration, including the Carnegie Institution for Science, Chilean National Time Allocation Committee (CNTAC) ratified researchers, the Gotham Participation Group, Harvard University, The Johns Hopkins University, L'Ecole polytechnique fédérale de Lausanne (EPFL), Leibniz-Institut f\"{u}r Astrophysik Potsdam (AIP), Max-Planck-Institut f\"{u}r Astronomie (MPIA Heidelberg), Max-Planck-Institut f\"{u}r Extraterrestrische Physik (MPE), Nanjing University, National Astronomical Observatories of China (NAOC), New Mexico State University, The Ohio State University, Pennsylvania State University, Smithsonian Astrophysical Observatory, Space Telescope Science Institute (STScI), the Stellar Astrophysics Participation Group, Universidad Nacional Autónoma de México, University of Arizona, University of Colorado Boulder, University of Illinois at Urbana-Champaign, University of Toronto, University of Utah, University of Virginia, and Yale University.

\bibliography{Paper2.bib}{}
\bibliographystyle{aasjournal}

\appendix
\section{The effect of gas pressure and $E_{peak}$ on Theoretical diagnostics for NLR of AGN}

\begin{figure*}[htb]
\epsscale{1.0}
\plotone{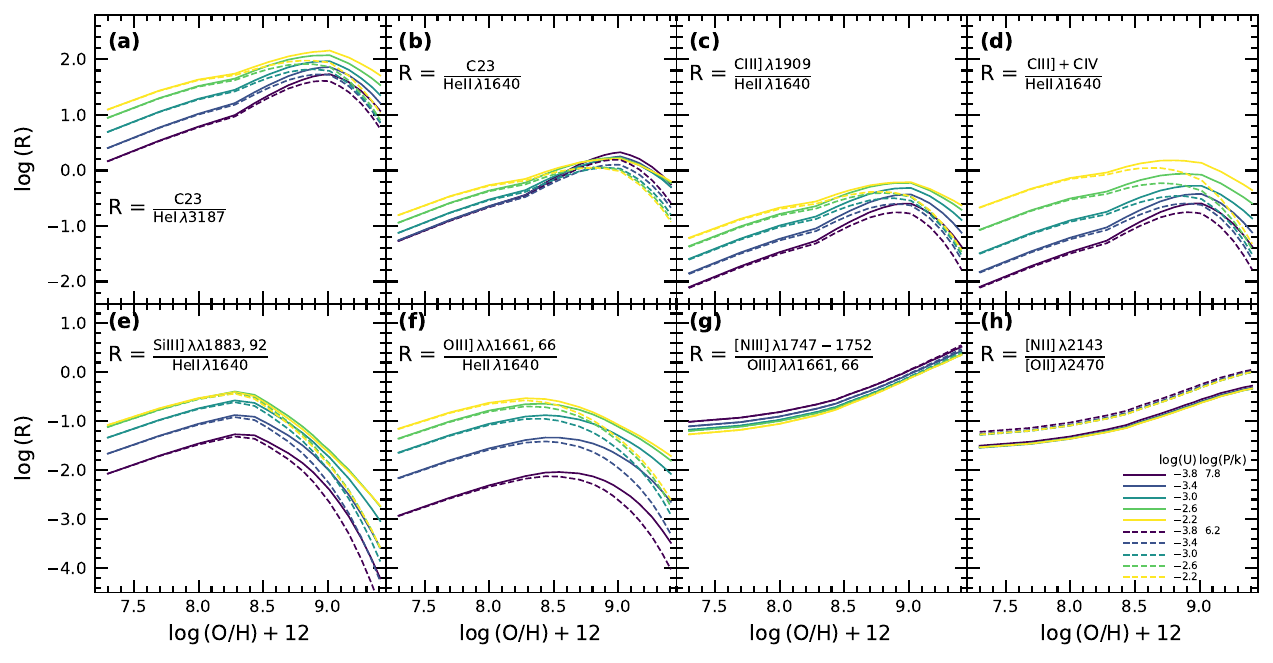}
\caption{{\color{black}The effect of gas pressure on UV metallicity diagnostics for the AGN narrow line regions as predicted by AGN models with $\log(E_{\rm peak}/\rm keV)=-1.25$.} As in Figure~\ref{fig:1}, but with line styles representing gas pressure ($\log(P/k)=7.8$ in solid lines and $\log(P/k)=6.2$ in dash lines) 
\label{fig:A1}}
\end{figure*}

\begin{figure*}[htb]
\epsscale{1.0}
\plotone{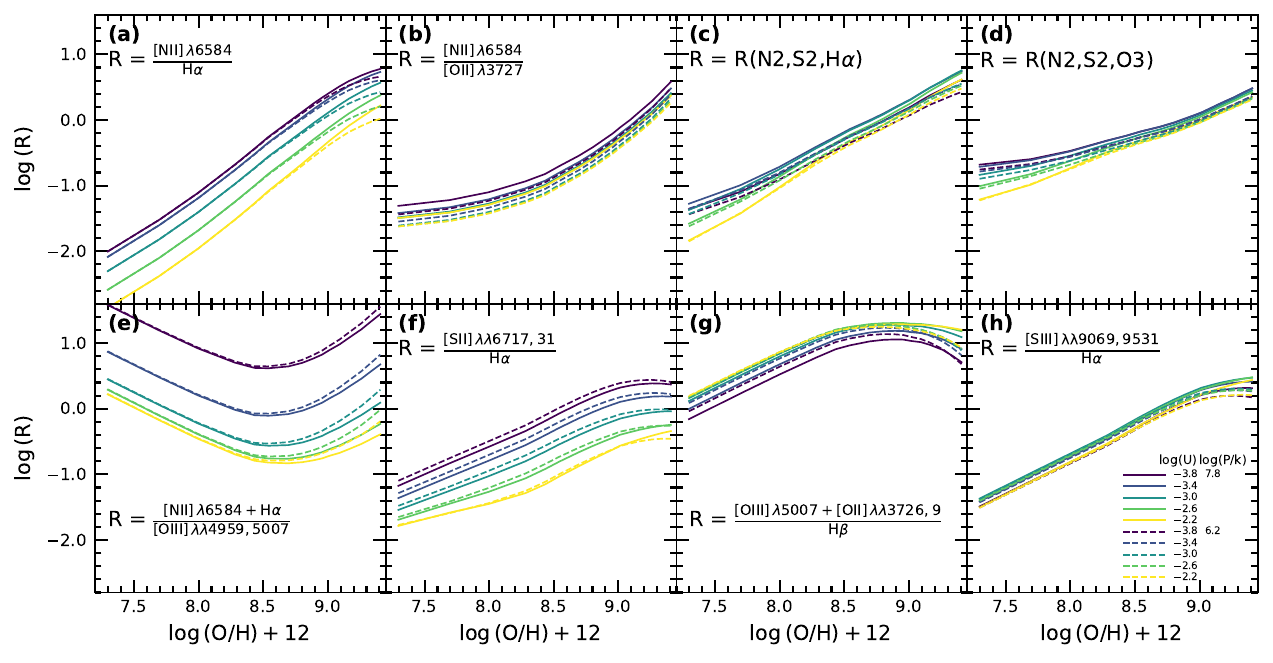}
\caption{{\color{black}The effect of gas pressure on optical metallicity diagnostics for the AGN narrow line regions as predicted by AGN models with $\log(E_{\rm peak}/\rm keV)=-1.25$.} As in Figure~\ref{fig:2}, but with line styles representing gas pressure ($\log(P/k)=7.8$ in solid lines and $\log(P/k)=6.2$ in dash lines) 
\label{fig:A2}}
\end{figure*}

%\begin{figure*}[htb]
%\epsscale{1.0}
%\plotone{ZIR_P.png}
%\caption{{\color{black}The effect of gas pressure on infrared metallicity diagnostics.} As in Figure~\ref{fig:3}, but with line styles representing gas pressure ($\log(P/k)=7.8$ in solid lines and $\log(P/k)=6.2$ in dash lines) 
%\label{fig:A3}}
%\end{figure*}

\begin{figure*}[htb]
\epsscale{1.0}
\plotone{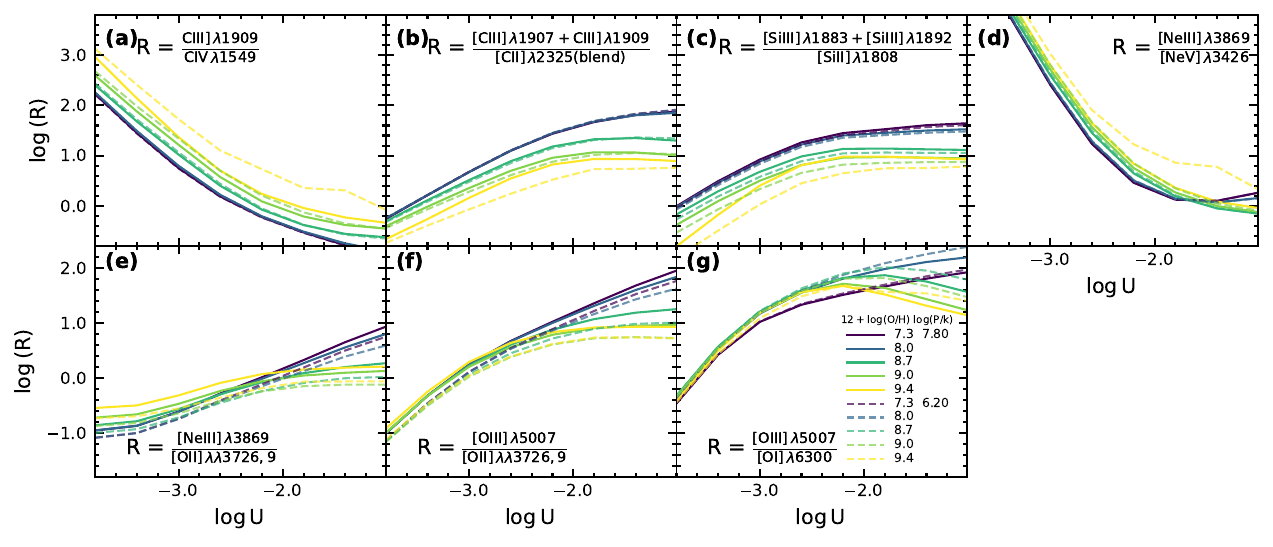}
\caption{{\color{black}The effect of gas pressure on UV and optical ionization parameter diagnostics for the AGN narrow line regions as predicted by AGN models with $\log(E_{\rm peak}/\rm keV)=-1.25$.} As in Figure~\ref{fig:4}, but with line styles representing gas pressure ($\log(P/k)=7.8$ in solid lines and $\log(P/k)=6.2$ in dash lines) 
\label{fig:A4}}
\end{figure*}

\begin{figure*}[htb]
\epsscale{1.0}
\plotone{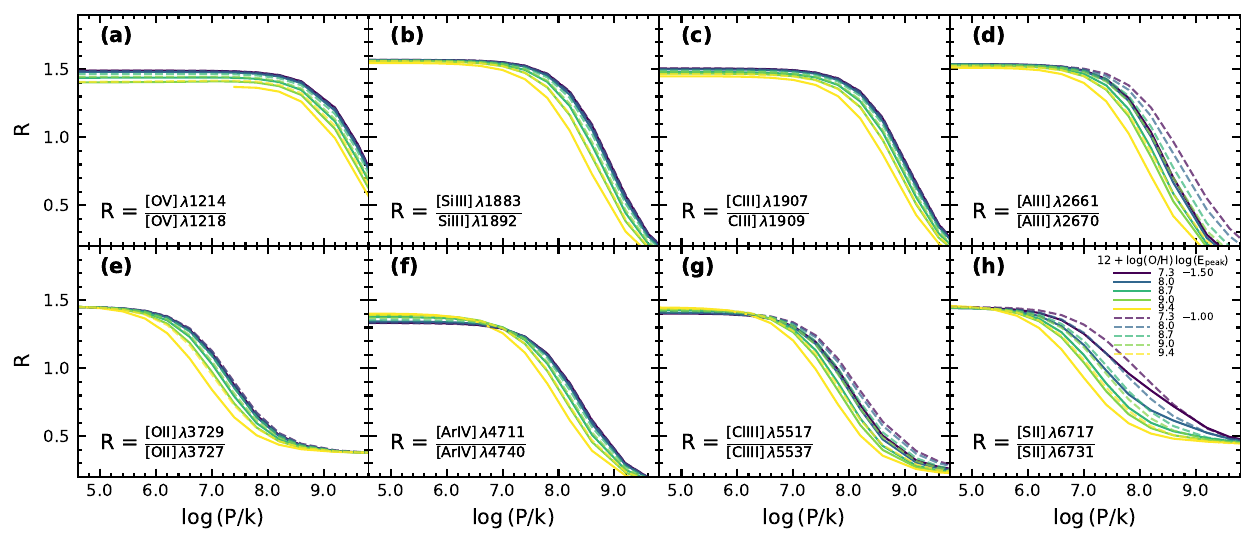}
\caption{{\color{black}The effect of $E_{peak}$ on UV and optical gas pressure diagnostics for the AGN narrow line regions as predicted by AGN models with $\log(U)=-3.0$.} As in Figure~\ref{fig:5}, but with line styles representing the peak energy in the AGN radiation field ($\log(E_{\rm peak}/\rm keV)=-1.0$ in solid lines and $\log(E_{\rm peak}/\rm keV)=-1.5$ in dash lines). 
\label{fig:A5}}
\end{figure*}

\begin{figure*}[htb]
\epsscale{1.0}
\plotone{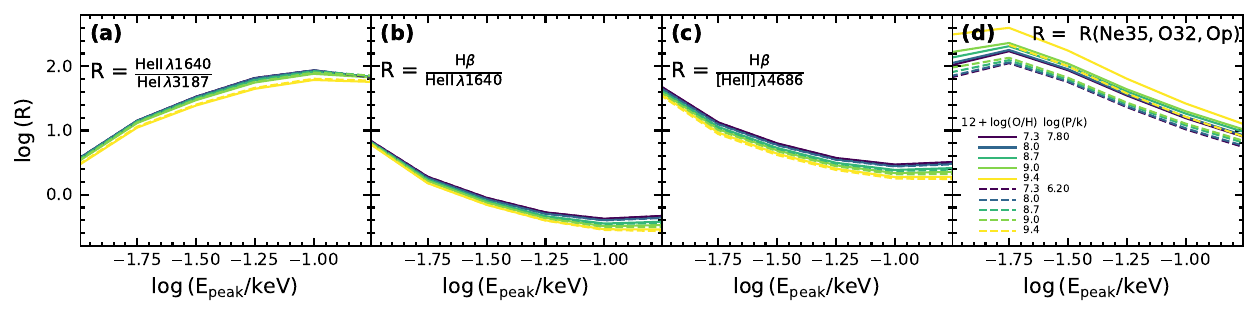}
\caption{{\color{black}The effect of gas pressure on UV and optical $E_{peak}$ diagnostics for the AGN narrow line regions as predicted by AGN models with $\log(U)=-3.0$.} As in Figure~\ref{fig:6}, but with line styles representing gas pressure ($\log(P/k)=7.8$ in solid lines and $\log(P/k)=6.2$ in dash lines) 
\label{fig:A6}}
\end{figure*}

\clearpage
\section{The effect of nitrogen scaling relation on AGN theoretical diagnostics for ionization parameter, gas pressure, and Epeak}

%\begin{figure*}[htb]
%\epsscale{1.0}
%\plotone{ZIR_NH.png}
%\caption{{\color{black}The effect of nitrogen scaling relation on infrared metallicity diagnostics.} As in Figure~\ref{fig:3}, but with line styles representing the nitrogen scaling relation used in the AGN models. (`NHlow' in solid lines and `NHhigh' in dash lines) 
%\label{fig:B3}}
%\end{figure*}

\begin{figure*}[htb]
\epsscale{1.0}
\plotone{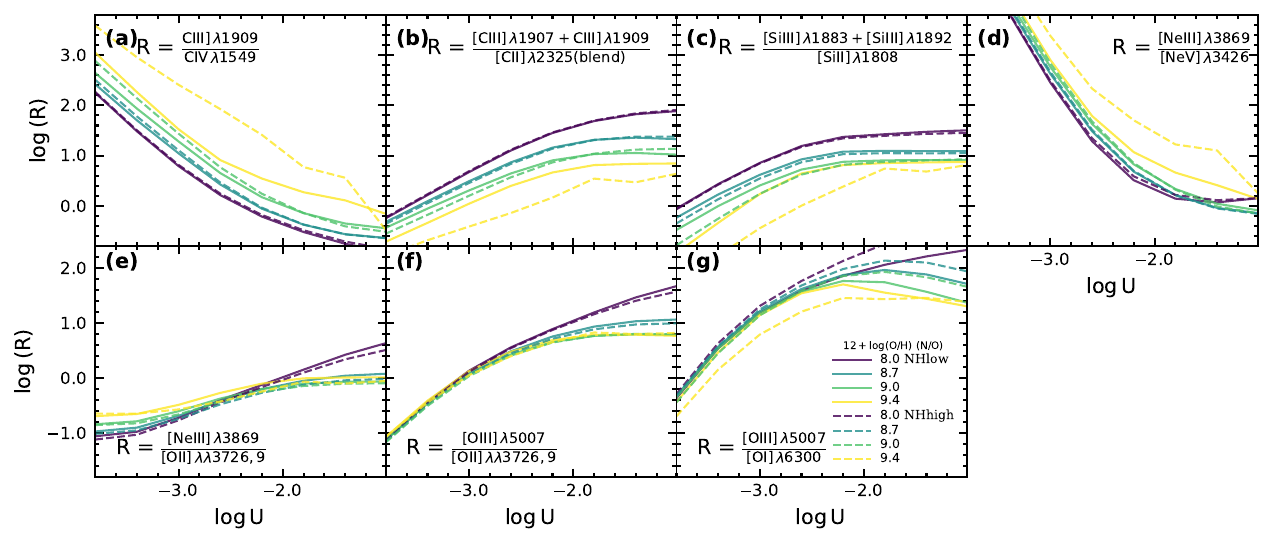}
\caption{{\color{black}The effect of nitrogen scaling relation on UV and optical ionization parameter diagnostics for the AGN narrow line regions as predicted by AGN models with $\log(P/k)=7.0$ and $\log(E_{\rm peak}/\rm keV)=-1.25$.} As in Figure~\ref{fig:4}, but with line styles representing the nitrogen scaling relation used in the AGN models. (`NHlow' in solid lines and `NHhigh' in dash lines)  
\label{fig:B4}}
\end{figure*}

\begin{figure*}[htb]
\epsscale{1.0}
\plotone{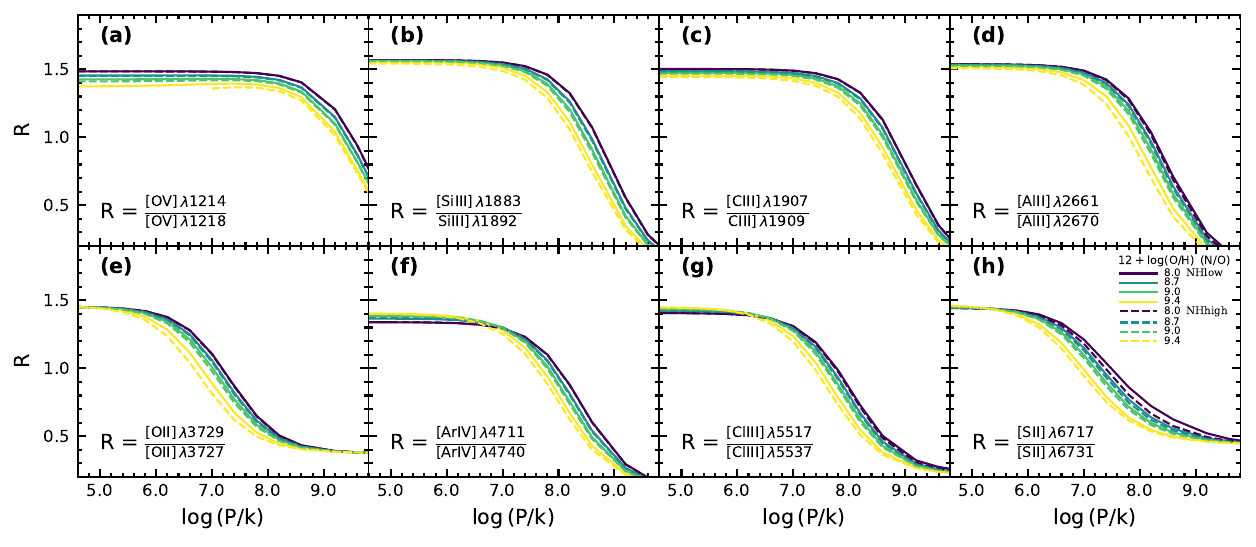}
\caption{{\color{black}The effect of nitrogen scaling relation on UV and optical gas pressure diagnostics for the AGN narrow line regions as predicted by AGN models with $\log(U)=-3.0$ and $\log(E_{\rm peak}/\rm keV)=-1.25$.} As in Figure~\ref{fig:5}, but with line styles representing the nitrogen scaling relation used in the AGN models. (`NHlow' in solid lines and `NHhigh' in dash lines)  
\label{fig:B5}}
\end{figure*}

\begin{figure*}[htb]
\epsscale{1.0}
\plotone{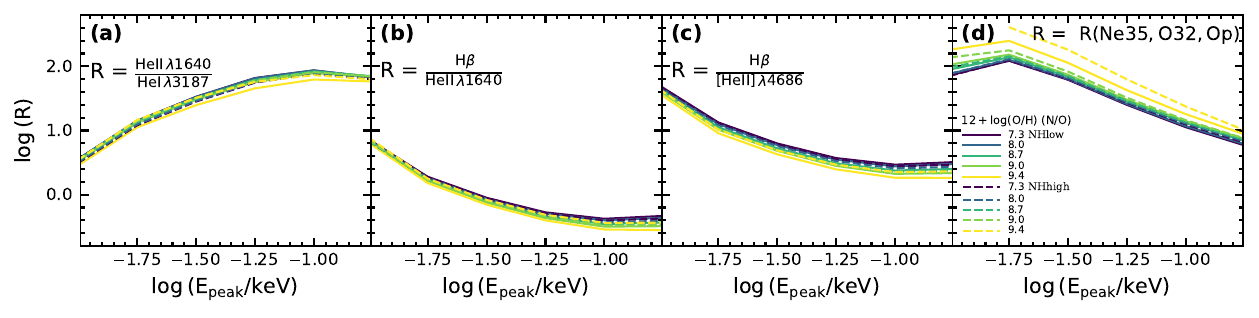}
\caption{{\color{black}The effect of nitrogen scaling relation on UV and optical $E_{peak}$ diagnostics for the AGN narrow line regions as predicted by AGN models with $\log(P/k)=7.0$ and $\log(U)=-3.0$.} As in Figure~\ref{fig:6}, but with line styles representing the nitrogen scaling relation used in the AGN models. (`NHlow' in solid lines and `NHhigh' in dash lines)  
\label{fig:B6}}
\end{figure*}

%first finished on Mar.21, 2024 T.T
%ready to submit on Oct.18, 2024
\end{document}